\documentclass[lettersize,journal]{IEEEtran}
\usepackage{amsmath,amsfonts}
\usepackage[noend]{algpseudocode}
\usepackage{algorithm}
\usepackage{array}
\usepackage[caption=false,font=normalsize,labelfont=sf,textfont=sf]{subfig}
\usepackage{textcomp}
\usepackage{stfloats}
\usepackage{url}
\usepackage{verbatim}
\usepackage{graphicx}
\usepackage{cite}
\usepackage{physics}
\usepackage{tikz}
\usetikzlibrary{quantikz}
\usepackage{braket}
\usepackage[bf]{caption} 

\usepackage{adjustbox}
\usepackage{algorithm,algpseudocode}

\usepackage{amsfonts}
\usepackage{amssymb}
\usepackage{amsmath}
\DeclareMathOperator*{\argmax}{argmax} 
\usepackage{lipsum}
\usepackage{wrapfig}
\usepackage{orcidlink}

\begin{document}

\title{A Quantum Fingerprinting Algorithm for Next Generation Cellular Positioning}

\author{Yousef Zook\orcidlink{0000-0003-1105-5464},~\IEEEmembership{Member,~IEEE}, Ahmed Shokry\orcidlink{0000-0003-3753-8886}, ~\IEEEmembership{Member, ~IEEE} and Moustafa Youssef\orcidlink{0000-0002-2063-4364}, ~\IEEEmembership{Fellow, ~IEEE}
\thanks{This work has been submitted to the IEEE for possible publication. Copyright may be transferred without notice, after which this version may no longer be accessible.}
\thanks{Yousef Zook is with the Alexandria University, Egypt (e-mail: es-yousif.mohamed14152@alexu.edu.eg). }
\thanks{Ahmed Shokry is with the American University in Cairo, Egypt (e-mail: ahmed.shokry@aucegypt.edu).}
\thanks{Moustafa Youssef is with the American University in Cairo and the University of New South Wales, Australia (e-mail: moustafa-youssef@aucegypt.edu).}}

\markboth{Journal of  IEEE JOURNAL ON SELECTED AREAS IN COMMUNICATIONS,~Vol.~xx, No.~x, x~2023}%
{Yousef Zook \MakeLowercase{\textit{et al.}}: Cosine Similarity-based Quantum Algorithm for Next Generation Cellular Positioning}


\maketitle

\begin{abstract}
The recent release of  
the third generation partnership project, Release 17,  calls for sub-meter cellular positioning accuracy with reduced latency in calculation.  
 To provide such high accuracy on a worldwide scale, leveraging the received signal strength (RSS) for positioning promises ubiquitous availability in the current and future equipment. 
RSS Fingerprint-based techniques have shown a great potential for providing high accuracy in both indoor and outdoor environments.  However, fingerprint-based positioning faces the challenge of providing a fast matching algorithm that can scale worldwide.

In this paper, we propose a cosine similarity-based quantum algorithm for enabling fingerprint-based high accuracy and worldwide positioning that can be integrated with the next generation of 5G and 6G networks and beyond. By entangling the test RSS vector with the fingerprint RSS vectors, the proposed quantum algorithm has a complexity that is exponentially better than its classical version as well as the state-of-the-art quantum fingerprint positioning systems, both in the storage space and the running time. 

We implement the proposed quantum algorithm and evaluate it in a cellular testbed on a real IBM quantum machine. Results show the exponential saving in both time and space for the proposed quantum algorithm while keeping the same positioning accuracy compared to the traditional classical fingerprinting techniques and the state-of-the-art quantum algorithms.
\end{abstract}


\section{Introduction}
\label{sec:introduction}

\IEEEPARstart{N}{owadays}, location determination services are crucial for many applications in both outdoor \cite{outdoor, shokry2018deeploc} and indoor \cite{youssef2015towards,fingerprint5GIndoor,crowdsourding5GIndoor} environments; such as emergency services, navigation, location-based analytics, among many others.

Supporting various positioning methods to provide accurate user equipment (UE)'s position has been one of the main features of the 3rd generation partnership project (3GPP) \cite{3gpp_rel17}. Release-17 further provides support for improved positioning in specific use cases such as factory automation by targeting sub-meter accuracy. In addition, Release-17 also introduces enhancements to latency reduction,  enabling positioning in time-critical use cases such as remote-control applications \cite{3gpp_rel17}.

Although different signals have been introduced for positioning \cite{5g_nr_doa, 5g_toa_doa, aoa_utdoa_5g_evaluation} such as round trip time, downlink and uplink time difference of arrival, and angle of arrival and departure, their current and future \textbf{ubiquitous}  deployment in cellular equipment is hard to be achieved. In contrast, the received signal strength (RSS) measurements are available in all cellular equipment to help in different  decisions, e.g., handoff. Therefore, RSS-based positioning techniques can provide a basis for ubiquitous cellular positioning on a worldwide scale for both indoor and outdoor environments.

Fingerprinting-based positioning is one of the mainstream technologies for RSS-based positioning \cite{ibrahim2011cellsense, gsm_indoor_2007_fingerprint,robust_loca_gsm_indoor_fingerprint,omnicells, energy_efficient_cellID_outdoor, metropolitan_GSM_positioning, youssef2005horus, bahl2000radar} due to its accuracy that can meet the recent Release-17 requirements. The fingerprint-based positioning technique is generally composed of two main phases: the offline phase and the online phase. In the offline phase, the signals received from the different base stations (BS) are scanned at different known locations in the environment. To construct the fingerprint, the received signal strength values (RSS's) are stored in a database along with the user’s location for each RSS value. Then, in the online phase, the online heard RSS at this time is matched with the collected fingerprint records in the database. Finally, the estimated location of the UE is the location in the fingerprint database that has the highest matching score with the current heard RSS.

The number of fingerprint records collected and the number of base stations affect the overall positioning accuracy: the higher the fingerprint locations and the BSs number are, the more precise the positioning will be\cite{bahl2000radar, youssef2005horus}. However, the time needed to match the heard RSS with the fingerprint data also significantly increases with the number of fingerprint records and BSs. Strictly speaking, the classical fingerprint-based positioning systems  (e.g. \cite{gsm_indoor_2007_fingerprint,robust_loca_gsm_indoor_fingerprint,omnicells, bahl2000radar, youssef2005horus, ibrahim2011cellsense}) need \textbf{$o(MN)$} space and their matching process runs in $o(MN)$, i.e. quadratic complexity, where $N$ is the number of BSs in the environment and $M$ is the number of locations in the fingerprint. This complexity hinders both the scalability and accuracy of the current positioning systems to be deployed on a worldwide scale. 

Recently, quantum fingerprinting positioning techniques have been proposed to overcome the classical techniques' limitations~\cite{quantum_lcn,quantum_qce,device_indp_q,quantum_vision,quantum_arx}. They can achieve \textbf{$o(M\log(N))$} time and space complexity, i,e. \textbf{sub-quadratic complexity}. In this paper, we propose a cosine similarity-based quantum algorithm that achieves \textbf{$o(\log(MN))$} time and space complexity, i.e. \textbf{sub-linear complexity}, providing a promising technique  that can scale to the huge number of BSs and fingerprinting locations for worldwide next generation cellular positioning. This is exponentially better than its classical counterpart in both the number of BSs ($N$) and the size of the fingerprint ($M$). Moreover, it is exponentially better than the current quantum positioning algorithms in the fingerprint size ($M$).

Evaluation of the proposed quantum algorithm in a real cellular testbed using a real IBM quantum machine shows that it can achieve the same accuracy as the classical techniques. This comes with a better than exponential reduction in time and space requirements, i.e. the time and space complexity for the proposed quantum algorithm is $o(\log(MN))$ compared to $o(MN)$ complexity for the classical counterpart, where $M$ is the number of fingerprint locations and $N$ is the number of BSs. Moreover, we compare the proposed algorithm with the state-of-the-art quantum positioning algorithms~\cite{quantum_lcn, quantum_qce, quantum_vision, quantum_arx, device_indp_q} and the classical techniques using a quantum machine simulator on a larger testbed. The results show that the proposed quantum algorithm can provide further exponential saving in the number of fingerprint locations, $M$, for both time and space. 

 
The rest of the paper is organized as follows: in Section~\ref{sec:related}, we discuss related work. Then in Section~\ref{sec:background}, we give a background on quantum computing. After that, we discuss the proposed quantum positioning algorithm and quantum circuit implementation in Section~\ref{sec:algorithm}. Then, we evaluate our system against other classical and quantum positioning systems in Section~\ref{sec:evaluation}. Finally, we conclude our work in Section~\ref{sec:conclusion}.

\section{Related Work}
\label{sec:related}
In this section, we discuss the different classical and quantum positioning algorithms. 

\subsection{Classical Positioning Systems}
Many classical algorithms have been developed for cellular-based positioning for both indoors~\cite{youssef2015towards, gsm_indoor_2007_fingerprint,robust_loca_gsm_indoor_fingerprint,omnicells} and outdoors~\cite{shokry2018deeploc,metropolitan_GSM_positioning,ibrahim2011cellsense,energy_efficient_cellID_outdoor, gsm_outdoor_m_youssef}. 

Generally, cellular-based positioning algorithms are based on the Cell-ID,  where the UE position is estimated using the serving cellular base station coordinates; using time-based and angle-based techniques, \cite{5g_nr_doa, 5g_toa_doa, aoa_utdoa_5g_evaluation}; or hybrid techniques \cite{ecid_5g_evaluation, cellid_ta_2005}. However, these methods depend on using the network infrastructure such as the BSs' coordinates or using external hardware (e.g. antenna array). 

On the other hand, using the received signal strength (RSS) values does not require extra hardware and is available from both the BS and the UE, making it more widely deployable than other methods~\cite{ibrahim2011cellsense, gsm_indoor_2007_fingerprint,robust_loca_gsm_indoor_fingerprint,omnicells, energy_efficient_cellID_outdoor, metropolitan_GSM_positioning}. 

Compared to propagation model-based systems, the majority of  RSS-based algorithms are based on fingerprint-matching techniques to provide higher accuracy~\cite{ibrahim2011cellsense,gsm_outdoor_m_youssef,youssef2005horus,bahl2000radar}. Those algorithms often use distance or similarity-based measurements like Euclidean distance, and cosine similarity \cite{bahl2000radar, beder2012fingerprinting, del2009efficient}. For example, \cite{metropolitan_GSM_positioning} provides a cellular-based positioning algorithm where a RSS vector is collected from different cellular towers for each location and stores the vectors in the fingerprint database. Then the unknown location is estimated by averaging the k-nearest fingerprint locations based on the Euclidean distance. 
Other systems use probabilistic techniques~\cite{ibrahim2011cellsense, youssef2005horus} where they store signal information distribution in the fingerprint during the offline phase, and try to estimate the most probable location in the online phase using this information. A common functionality between these different techniques is the need for matching the UE's RSS measurements in the online phase with the fingerprint data collected in the offline phase, which takes $o(MN)$ in both time and space. 

In contrast, this paper proposes a cosine similarity fingerprinting \textit{quantum algorithm} that takes $o(\log(MN))$ time and space.

\subsection{Quantum Positioning Systems}
Quantum algorithms have shown exponential speed gain in different areas recently. They leverage quantum properties to enhance the classical algorithms' time and space. Examples include the Grover's algorithm \cite{grover} that provides an unstructured search technique in $o(\sqrt{n})$ instead of $o(n)$ provided by the classical version.  Similarly, the well-known Shor's algorithm uses quantum computing to efficiently factor large numbers in polynomial time, raising the possibility of breaking the commonly-used RSA encryption technique \cite{shor}.

Quantum algorithms for positioning have gained momentum recently~\cite{device_indp_q, quantum_arx, quantum_vision, quantum_lcn, quantum_qce, qps, q_loc_survey}. In~\cite{qps}, the author proposes a quantum version of the classical GPS that can provide a user with his/her coordinates. However, it leverages the quantum entanglement for clock synchronization. In~\cite{q_loc_survey}, the authors discuss using quantum inertial sensors to locate the user equipment. However, the previous quantum techniques require special hardware and synchronization. In contrast, the authors of \cite{device_indp_q}, propose a device-independent quantum fingerprint-matching algorithm that can work for heterogeneous standard WiFi devices with space and running time complexity of $o(M\log(N))$. In~\cite{quantum_vision}, the authors discuss the challenges and opportunities of using quantum computing in positioning techniques. In \cite{quantum_lcn, quantum_qce, quantum_arx}, the authors propose a cosine similarity-based quantum algorithm for a positioning system that is exponentially faster than its classical version in the dimension of the number of BSs in the space and time complexity ($o(M\log(N))$).

Unlike the previous quantum positioning algorithms, the proposed quantum algorithm pushes the space and time complexity to be better than exponentially more efficient than the state-of-the-art quantum algorithms with a time and space complexity of $o(\log(MN))$.


\section{Quantum Computing Background and Notation}
\label{sec:background}
In this section, we will provide a background on the quantum computing basic concepts that will be used in our algorithm. 

Quantum computing \cite{qbible} is an area of computing based on quantum mechanics theories. It mixes three fields: mathematics, physics and computer science. The basic unit of processing in quantum computing is the \textit{qubit}. Qubits are the quantum version of the classical bits that are used to store information. Similar to classical registers, quantum registers are used to hold a group of qubits. A qubit can be presented as a photon polarization or an electron spin, which allows a qubit to be in any state of an infinite number of states between zero and one. This phenomenon is called \textit{quantum superposition}. It can be leveraged along with other quantum properties to speed up classical computations. Mathematically, the state of a qubit can be represented  as a vector using the \textit{Dirac Notation} \cite{dirac_paper}. Ket ($\ket{.}$) is the Dirac's name for a \textit{column} vector, and it is used to represent a qubit state $\ket{\phi}$ as vector of sums of the basis states ${\ket{0} = \begin{bmatrix} 1 \\ 0\end{bmatrix}, \ket{1}= \begin{bmatrix} 0 \\ 1\end{bmatrix}}$ as $\ket{\phi} = \gamma\ket{0} + \eta\ket{1} = \begin{bmatrix} \gamma \\ \eta\end{bmatrix}$, where $\gamma$ and $\eta$ are complex numbers representing the probability amplitudes of measuring the qubit with value 0 and value 1 respectively\cite{qbible}, i.e. $\abs{\gamma}^2+\abs{\eta}^2=1$.
On the other hand, a row vector is represented as a Bra ($\bra{.}$), for example, if we want to represent the transpose of ket $\ket{\phi}$, we can write it as $\bra{\phi}=\begin{bmatrix} \gamma & \eta \end{bmatrix}$. The dot product of two vectors can be written as a bra-ket in Dirac's notation, e.g.,
 $\braket{\psi|\phi}$. For $n$ qubits, we can use the tensor product symbol ($^\otimes$). For example, we use $\ket{0}^{\otimes n}$ to  mean $n$ qubits in the state $\ket{0}$.

\begin{figure}[t!]
	\centering
\begin{quantikz}
	\ket{0} & \gate{X} &  \gate{H} & \meter{} &  \cw  \rstick[style={xshift=-10cm}]{$ \left\{ \begin{array}{rl}
			0 & \text{with prob. 0.5} \\
			1 & \text{with prob. 0.5} 
		\end{array} \right.$}
\end{quantikz}	
	\caption{A quantum circuit that inverts a single qubit and puts it into a superposition state. The $X$ block represents the NOT gate which inverts the qubit state ($\ket{0}\rightarrow\ket{1}$), the $H$ block represents the Hadamard gate which changes the qubit state $\ket{1}$ into an equal superposition state ($\ket{1}\rightarrow \frac{1}{\sqrt{2}}(\ket{0}-\ket{1}$)), and the third block represents the measurement. Single lines carry quantum information while double lines carry classical information.}
	\label{fig:simplecircuitNotHadamard}
\end{figure}
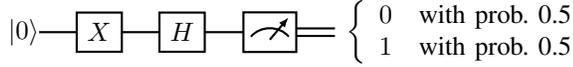
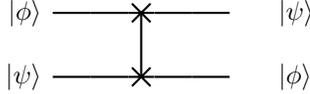
\begin{figure}[t!]
	\centering
	\begin{quantikz}
		\lstick{$\ket{\phi}$}  & \qw  & \swap{1} & \qw & \qw & \rstick{$\ket{\psi}$} \\ 
            \lstick{$\ket{\psi}$}  & \qw  & \targX{} & \qw & \qw & \rstick{$\ket{\phi}$} \\ 	
	\end{quantikz}	
	\caption{The Swap gate which exchanges the states of two qubits.}
	\label{fig:swap}
\end{figure}

A quantum algorithm is represented as a quantum circuit, composed of \textit{quantum gates}. Quantum gates are similar to classical gates but are applied on qubits and are represented mathematically as unitary matrices. For example, the NOT gate (X-gate) can be represented as a unitary matrix $\textrm{NOT} = \begin{bmatrix} 0 & 1 \\ 1 & 0\end{bmatrix}$. Hence, for a state $\ket{\phi} = \gamma\ket{0} + \eta\ket{1}$, applying the NOT operation leads to the state: $\textrm{NOT} \ket{\phi} = \eta\ket{0} + \gamma\ket{1}$. Another example is the Hadamard gate (H-gate) which produces an equal superposition state of $\ket{0}$ and $\ket{1}$ when it acts on a single qubit in either of the basis states, i.e. it maps the basis states $\ket{0}$ to $\frac{1}{\sqrt{2}}(\ket{0}+\ket{1})$, and $\ket{1}$ to $\frac{1}{\sqrt{2}}(\ket{0}-\ket{1})$. The H gate can be represented as $ \frac{1}{\sqrt{2}}\begin{bmatrix} 1 & 1 \\ 1 & -1\end{bmatrix}$. Figure~\ref{fig:simplecircuitNotHadamard} shows an example of a quantum circuit with a single qubit.  
Similarly, gates can also be applied to multiple qubits. For example,  the SWAP  gate  exchanges the quantum state of two qubits as shown in Figure~\ref{fig:swap}. 



\begin{figure}[t!]
	\centering
	\begin{quantikz}
		\lstick{$\ket{control}$} &\qw & \ctrl{1} & \qw & \qw\\ 
		\lstick{$\ket{target}$}  & \qw  & \targ{} & \qw & \qw \\	
	\end{quantikz}	
	\caption{An example of a controlled gate. Qubit ``$control$'' controls the NOT operation on another qubit, ``$target$''. The target qubit will be negated $\;\Longleftrightarrow\;$ the control is measured to be 1.}
	\label{fig:controlledNot}
\end{figure}
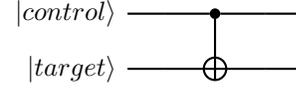



\begin{figure}[!t]
	\centerline
	{\includegraphics[width=0.55\textwidth]{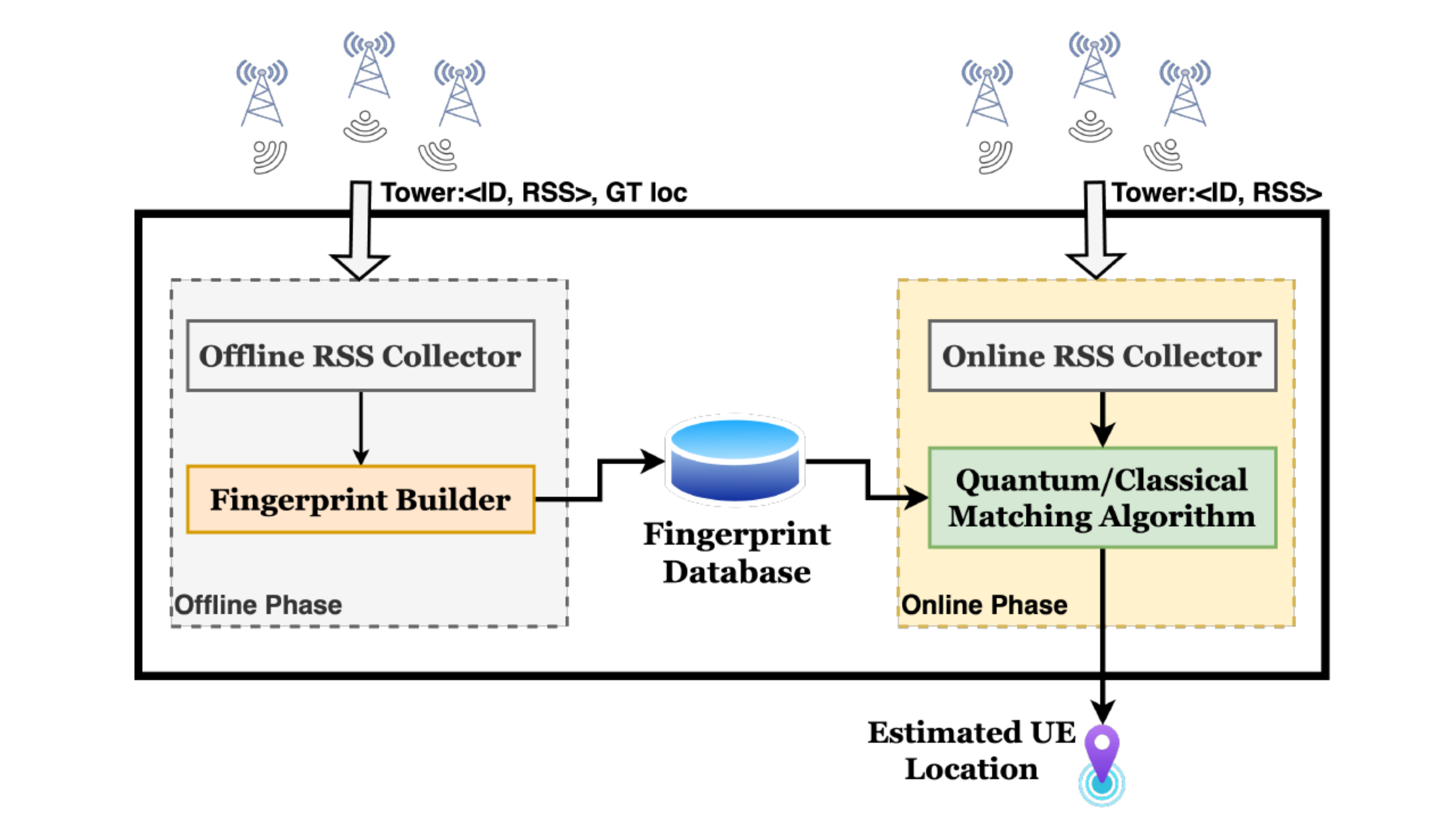}}
	\caption{General system architecture for the fingerprint-based positioning technique.}
	\label{fig:fingerprint_arch}
\end{figure}

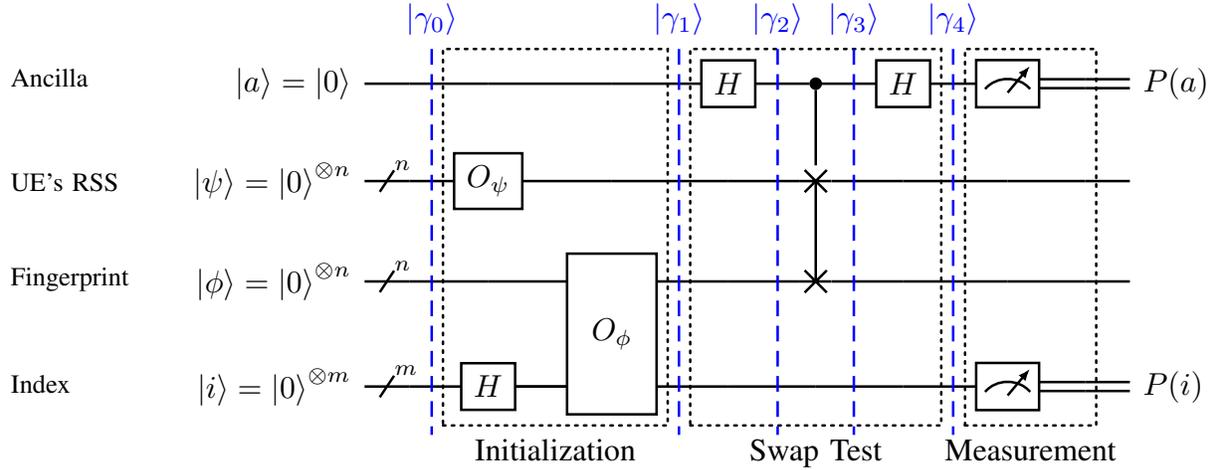
\begin{figure*}[t!]
	\centering
	\begin{adjustbox}{width=0.9\textwidth}
		\begin{tikzpicture}[row sep=0.1cm]
			\node[text width=1.5cm] at  (-8.5,  2.1){Ancilla};
			\node[text width=1.5cm] at  (-8.5,  0.7){UE's RSS};
			\node[text width=1.5cm] at  (-8.5, -0.6){Fingerprint};
			\node[text width=1.5cm] at  (-8.5, -2){Index};
					\node[scale=1.2] {
	\begin{quantikz}
		\lstick{$\ket{a} = \ket{0}$} & \qw\slice[style={color=blue},label style={color=blue}]{$\ket{\gamma_0}$} & \qw \gategroup[wires=4, steps=2, style={dotted, cap=round, inner sep=0pt}, label style={label position=below, yshift=-0.5cm}]{Initialization} & \qw \slice[style={color=blue},label style={color=blue}]{$\ket{\gamma_1}$} & \gate{H} \gategroup[wires=4, steps=3, style={dotted, cap=round, inner sep=0pt}, label style={label position=below, yshift=-0.5cm}]{Swap Test}\slice[style={color=blue},label style={color=blue}]{$\ket{\gamma_2}$} & \ctrl{1}\slice[style={color=blue},label style={color=blue}]{$\ket{\gamma_3}$} & \gate{H}\slice[style={color=blue},label style={color=blue}]{$\ket{\gamma_4}$} & \meter{} \gategroup[wires=4, steps=2, style={dotted, cap=round, inner sep=0pt}, label style={label position=below, yshift=-0.5cm}]{Measurement} & \cw & \cw \rstick{$P(a)$}\\
		\lstick{$\ket{\psi} = \ket{0}^{\otimes n}$}  & \qwbundle{n} & \gate{O_\psi} & \qw &  \qw & \swap{1} & \qw & \qw &  \qw &  \qw \\
		\lstick{$\ket{\phi} = \ket{0}^{\otimes n}$}  & \qwbundle{n} & \qw & \gate[wires=2][1cm]{O_\phi} & \qw & \targX{} & \qw & \qw & \qw & \qw\\	
		\lstick{$\ket{i} = \ket{0}^{\otimes m}$}  & \qwbundle{m} & \gate{H} & \qw & \qw & \qw &  \qw & \meter{} & \cw & \cw \rstick{$P(i)$}						
	\end{quantikz}
                    };
		\end{tikzpicture}
	\end{adjustbox}
	\caption{The quantum circuit to calculate the  cosine similarity between UE's RSS sample vector $\ket{\psi}$, and all fingerprint records in parallel $\ket{\phi}$. $\ket{\gamma_i}$ represents the joint system state at different positions in the circuit.}
	\label{fig:main_circuit}
\end{figure*}

\textit{Quantum entanglement} between qubits means that there is a dependency between them. For example, if we have two qubits and one qubit is measured and collapses to a specific state, then the other one will immediately jump to a certain state depending on the measured value of the first one. This can be achieved, e.g., using multi-qubit gates. An example of these gates is the Controlled NOT (CNOT) gate shown in Figure~\ref{fig:controlledNot}, where the control qubit controls the NOT operation on the target qubit. This entanglement phenomenon is used in different quantum applications like quantum teleportation \cite{teleportation} and superdense coding \cite{superdense}. We use the quantum entanglement phenomenon to add dependency between the collected cellular RSS vectors at known locations (i.e. the fingerprint) and the UE's RSS vector at an unknown location.

\textit{Quantum interference} refers to changing the probability amplitudes of a certain qubit using quantum gates. 
We use this to bias the probabilities of certain qubits to reflect the cosine similarity between the UE's RSS at an unknown location and the fingerprint.


\section{The Quantum Positioning Algorithm}
\label{sec:algorithm}
In this section, we discuss the details of the proposed quantum cellular positioning algorithm. The basic idea behind the proposed algorithm is to calculate the quantum cosine similarity between the online RSS vector at an unknown location and each RSS vector in the cellular fingerprint. This can be achieved by putting the fingerprint samples into a superposition quantum state and entangling them with the UE's RSS vector by applying a sequence of quantum gates. 

We start by discussing the quantum circuit for positioning. Then, we explain the implementation details of the circuit. Finally, we give a numerical example that explains the steps of the proposed algorithm.

\subsection{Positioning Algorithm}
The proposed quantum algorithm works in two phases: the
offline quantum fingerprint building phase and the online user
tracking phase as shown in Figure~\ref{fig:fingerprint_arch}. 

In the offline phase, we collect the fingerprint RSS data at known ground-truth (GT) locations. Each fingerprint sample contains the RSS vector from different BSs, e.g. cellular towers, in the environment and the location where the RSS vector is collected. In the online phase, the user's current location is queried based on the signals heard by the user equipment (UE). The UE scans the set of RSS from the different base stations in the environment and passes it along with the fingerprint to the quantum cosine similarity algorithm to find the cosine similarities between the current UE's RSS and the collected fingerprint data at each location. Finally, the fingerprint location with the highest similarity score is returned as the estimated location.

Without loss of generality, assume that we have $N$ base stations that can be heard at $M$ fingerprint locations. Also, assume that the online normalized RSS vector is $\psi$ and the offline normalized RSS vector at location $j$ is $\phi_j$, where $j \in {\{0,..,M-1}\}$. The cosine similarity between $\psi$ and $\phi_j$ for each $j$ is:

\begin{equation}
\label{eq:cosine_eq}
    \cos ({\bf \psi},{\bf \phi_j})= \big|\braket{\bf \psi | \bf \phi_j}\big| = \sum_{i=0}^{N-1}{{\bf \psi}_j{\bf \phi_j}_i}
\end{equation}

 

The quantum circuit  in Figure~\ref{fig:main_circuit} calculates the cosine similarity between $\psi$ and \textbf{all $\boldsymbol \phi_j$'s in parallel}. It consists of three stages: the Initialization, Swap Test, and Measurement stages. 

\subsubsection{The Initialization Stage} The input to the circuit is four quantum registers: a single ancilla qubit, an $n$-qubits register for encoding the UE's collected RSS sample $\ket{\psi}$, another $n$-qubits register to encode the fingerprint data $\ket{\phi}$ at each fingerprint location, and finally an $m$-qubits register to reflect the fingerprint location index $\ket{i}$. Initially, all registers are in the following state,

\begin{equation}
\label{eq:initial}
    \ket{\gamma_0} = \ket{0}\ket{0}^{\otimes n}\ket{0}^{\otimes n}\ket{0}^{\otimes m}
\end{equation}
Where $n=\log(N)$ and $ m=\log(M)$.
The first step during the initialization is to convert the classical RSS data to quantum data. To do this for the current UE's RSS vector, we apply the oracle $O_\psi$ to the $n$-qubits register in the zero state $\ket{0}^{\otimes n}$, where $O_\psi$ is a gate/circuit that converts the register to the state $\ket{\psi}$ as shown in Equation~\ref{eq:init_psi_eq}. We explain how to implement $O_\psi$ 
 later in this section.

\begin{equation}
\label{eq:init_psi_eq}
    O_\psi \ket{0}^{\otimes n} = \ket{\psi}
\end{equation}

Similarly, we encode the RSS vector at each  fingerprint location $j$ with its index using the Hadamard gate and oracle $O_\phi$. The Hadamard gate converts the index register $\ket{i}$ to a superposition state with equal probabilities as shown in Equation~\ref{eq:hadamard_index}. 

\begin{equation}
\label{eq:hadamard_index}
    H \ket{0}^{\otimes m} = \frac{1}{\sqrt{M}}\sum_{j=0}^{M-1} \ket{j} = \ket{i}
\end{equation}

The oracle $O_\phi$ is another gate/circuit that converts qubits to $\ket{\phi_j}$ entangled with the index register as

\begin{equation}
\label{eq:oracle_phi}
    O_\phi \ket{0}^{\otimes n} \ket{i} = \frac{1}{\sqrt{M}}\sum_{j=0}^{M-1} \ket{\phi_j} \ket{j}
\end{equation}
where $j$ represents numbers from 0 to $M-1$ (index of fingerprint data at location $j$ with size $M$). We give the details of the $O_\phi$ oracle later in this Section.

After the initialization stage, the system becomes in the following state, 

\begin{equation}
\label{eq:gamma_1}
     \ket{\gamma_1} = \frac{1}{\sqrt{M}} \sum_{j=0}^{M-1} \ket{0}\ket{\psi}\ket{\phi_j}\ket{j}
\end{equation}

\subsubsection{The Swap Test Stage} The next step is to do a swap test \cite{swap_paper}. The goal is to entangle the ancilla qubit, the UE's RSS sample register and the fingerprint register to calculate the similarity score in parallel for all fingerprint locations.  We start by applying the Hadamard gate to the ancilla qubit, which leads to the following state

\begin{equation}
\label{eq:gamma_2}
     \ket{\gamma_2} = \frac{1}{\sqrt{2M}} \sum_{j=0}^{M-1} (\ket{0} + \ket{1})\ket{\psi}\ket{\phi_j}\ket{j}
\end{equation}

Then, the ancilla is entangled with $\ket{\psi}$ and $\ket{\phi}$ registers using a controlled swap gate which leads to the following state,
 
\begin{equation}
\label{eq:gamma_3}
\begin{aligned}
     \ket{\gamma_3}=\frac{1}{\sqrt{2M}} \sum_{j=0}^{M-1} (\ket{0}\ket{\psi}\ket{\phi_j}\ket{j}
    + \ket{1}\ket{\phi_j}\ket{\psi}\ket{j})
\end{aligned}
\end{equation}

The last step in the swap test is applying the Hadamard gate again to the ancilla qubit which leads to the following final state,

\begin{equation}
\label{eq:gamma_bar}
\begin{aligned}
     \ket{\gamma_4} = \frac{1}{2\sqrt{M}} \sum_{j=0}^{M-1} (\ket{0}[\ket{\psi}\ket{\phi_j} + \ket{\phi_j}\ket{\psi}]\\
    + \ket{1}[\ket{\psi}\ket{\phi_j} - \ket{\phi_j}\ket{\psi}])\ket{j}
\end{aligned}
\end{equation}

\subsubsection{The Measurement Stage} The final stage is the measurement stage, where we measure the ancilla qubit conditioned on the index register being in state $\ket{j}$, i.e. $p(a|i=j)$, where $j$ represents the index number in the set $\{0,..,M-1\}$. This probability, $p(a|i=j)$, is a function of the required cosine similarity between the UE's RSS sample and fingerprint sample $j$. 




In particular, to find this probability $p(a|i = j)$; we measure the index register first (since the probability is conditioned on the index value). This moves the state of the unmeasured quantum system to (see Appendix~\ref{appendix:A_1} for details):


\begin{equation}
\label{eq:gamma_tilde}
\begin{aligned}
     \frac{1}{2}(\ket{0}[\ket{\psi}\ket{\phi_j} + \ket{\phi_j}\ket{\psi}] + \ket{1}[\ket{\psi}\ket{\phi_j} - \ket{\phi_j}\ket{\psi}])
\end{aligned}
\end{equation}

Then, we can calculate the probability that the ancilla is zero given that the index is $j$ as follows (this can be obtained by normalizing the $\psi$ and $\phi$ states, see Appendix~\ref{appendix:A_2} for details):

\begin{equation}
\label{eq:p_a_given_i}
\begin{aligned}
    p(a=0|i=j) = \big(\frac{1}{2} \times \sqrt{2+2\big|\braket{\psi|\phi_j}\big|^2}\big)^2 \\
    = \frac{1}{2} + \frac{1}{2} \big|\braket{\psi|\phi_j}\big|^2
\end{aligned}
\end{equation}

Hence, the cosine similarity can be obtained as:
\begin{equation}
\label{eq:cos_final}
\begin{aligned}
    cos(\psi,\phi_j) = \big|\braket{\psi|\phi_j}\big| = \sqrt{2\times p(a=0|i=j) - 1}
\end{aligned}
\end{equation}
where $p(a=0|i=j)$ is the conditional probability of measuring the ancilla qubit to be $0$ conditioned on that the index register $i$ is equal to $j$, where $j \in \{0,..,M-1\}$, and $M$ is the size of the fingerprint.


The probability $p(a=0|i=j)$ can be found by running the circuit for $K$ times (the number of shots in the quantum terminology) and calculating 

\begin{equation}
\label{eq:probability_by_counting}
\begin{aligned}
    p(a=0|i=j) = \frac{\text{count}(a=0 \cap i=j)}{\text{count}(i=j)}
\end{aligned}
\end{equation}

The estimated location is the location of the fingerprint sample $j$ with the highest cosine similarity, i.e. $j=\argmax_j (cos(\psi,\phi_j))$. Since this cosine similarity 
 is directly proportional to $\text{count}(a=0 \cap i=j)$ (see Appendix~\ref{appendix:B} for details), then the estimated location is the location of the fingerprint sample that has the maximum count of measuring the ancilla qubit output as $0$.

Algorithm~\ref{main_alg} summarizes the proposed quantum positioning algorithm. 


\begin{algorithm} 
\caption{$o(\log MN)$ Quantum Positioning}
\label{main_alg}
\begin{algorithmic}[1]
\Require
\Statex - Two $n$-qubits quantum registers $\ket{\phi}$ and  $\ket{\psi}$,  where $\ket{\phi}$ is used to hold the fingerprint data, and $\ket{\psi}$ is used to hold the test sample, $n = \log(N)$, $N$ is the number of BSs.
\Statex - An ancilla qubit, $\ket{a} = \ket{0}$. 
\Statex - A quantum register with $m$ qubits, $\ket{i}$, that holds the index value. $m = \log(M)$, where $M$ is the number of fingerprint samples.
\Statex - Number of shots $K$.
\Statex - $\text{gt\_loc}[]$: the ground truth locations from the fingerprint data.
\Statex
\Ensure The user equipment position.
\Statex
\State {$\textrm{counts}[]$ $\gets$ zeros($M$) \Comment\small{Array to count $\ket{a}=0$ at each fingerprint index}}
\State {$\textrm{max\_count}$ $\gets$ 0} 
\State {$\textrm{max\_index}$ $\gets$ 0} 
\For{$k \gets 1$ to $K$}         

    /* Initialization stage */
    \State {$\ket{\psi} \gets$ Apply $O_{\psi}(\ket{0}^{\otimes n})$   \Comment\small{UE's sample initialization}}
    \State {$\ket{i} \gets$ Apply $H(\ket{0}^{\otimes m})$ \Comment\small{Index initialization}}
    \State {$\ket{\phi} \gets$ Apply $O_{\phi}(\ket{0}^{\otimes n},\ket{i})$\Comment\small{Fingerprint initialization}}

    /* Swap Test stage */
    \State {Apply $H(\ket{a})$}
    \State {Apply $\text{CSWAP}(\ket{a},\ket{\psi},\ket{\phi})$}
    \State {Apply $H(\ket{a})$}

    /* Measurement stage */
    \State {$j \gets \textrm{measure}(\ket{i})$ \Comment\small{j is the measured value of the index register ($\ket{i}$)}}
    \If{$\textrm{measure}(\ket{a}) = 0$}
        \State {$\textrm{counts}[j] \gets \textrm{counts}[j]+1$ \Comment\small{count($a=0 \cap i=j$)}}
    \EndIf
    \If{$\textrm{counts}[j] > \textrm{max\_count}$}
        \State {$\textrm{max\_count} \gets \textrm{counts}[j]$}
        \State {$\textrm{max\_index} \gets j$}
    \EndIf
\EndFor
\State \Return {$\text{gt\_loc}[\text{max\_index}]$}
\end{algorithmic}
\end{algorithm}

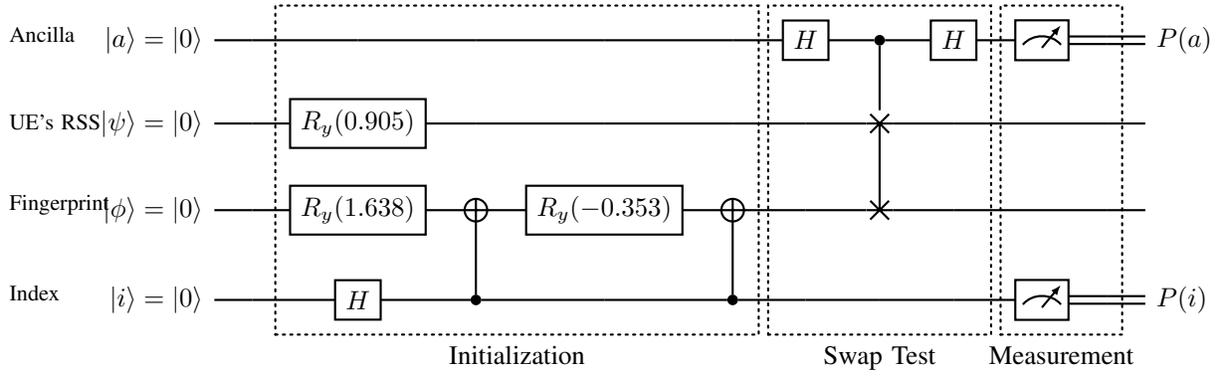
\begin{figure*}[t!]
	\centering
	\begin{adjustbox}{width=0.9\textwidth}
		\begin{tikzpicture}[row sep=0.1cm]
			\node[text width=1.5cm] at  (-9.5,  2.5){Ancilla};
			\node[text width=1.5cm] at  (-9.5,  1.1){UE's RSS};
			\node[text width=1.5cm] at  (-9.5, -0.2){Fingerprint};
			\node[text width=1.5cm] at  (-9.5, -1.6){Index};
					\node[scale=1.2] {
	\begin{quantikz}
		\lstick{$\ket{a} = \ket{0}$} & \qw & \qw \gategroup[wires=4, steps=4, style={dotted, cap=round, inner sep=2pt}, label style={label position=below, yshift=-0.5cm}]{Initialization} & \qw & \qw & \qw & \gate{H} \gategroup[wires=4, steps=3, style={dotted, cap=round, inner sep=2pt}, label style={label position=below, yshift=-0.5cm}]{Swap Test} & \ctrl{1} & \gate{H} & \meter{} \gategroup[wires=4, steps=2, style={dotted, cap=round, inner sep=2pt}, label style={label position=below, yshift=-0.5cm}]{Measurement} & \cw & \cw \rstick{$P(a)$}\\
		\lstick{$\ket{\psi} = \ket{0}$}  & \qw & \gate{R_y(0.905)} & \qw & \qw & \qw &  \qw & \swap{1} & \qw & \qw &  \qw &  \qw \\
		\lstick{$\ket{\phi} = \ket{0}$}  & \qw & \gate{R_y(1.638)} &  \targ{} & \gate{R_y(-0.353)} &  \targ{} & \qw & \targX{} & \qw & \qw & \qw & \qw\\   
		\lstick{$\ket{i} = \ket{0}$}  & \qw & \gate{H} & \ctrl{-1} & \qw & \ctrl{-1} & \qw  & \qw &  \qw & \meter{} & \cw & \cw \rstick{$P(i)$} 
	\end{quantikz}
                    };
		\end{tikzpicture}
	\end{adjustbox}
	\caption{ Example quantum circuit for cosine similarity.}
	\label{fig:example_circuit}
\end{figure*}

\subsection{Quantum Circuit Implementation}
\label{sec:impel}

The quantum circuit in Figure~\ref{fig:main_circuit} contains two oracles. The first oracle, $O_\psi$, is used to initialize the qubit register with the UE's current  RSS values as shown in Equation~\ref{eq:init_psi_eq}. Specifically, given a classical RSS vector $\begin{bmatrix} \alpha_0 & \alpha_1 & ... & \alpha_{N-1}\end{bmatrix}$, the goal of the oracle is to convert the vector $\begin{bmatrix} 0 & 0 & ... & 0\end{bmatrix}$ to $\begin{bmatrix} \alpha_0 & \alpha_1 & ... & \alpha_{N-1}\end{bmatrix}$, i.e., encode the classical RSS vector into the probability amplitudes of the $n$-qubits register $\ket{\psi}$.  This amplitude encoding can be done generally using quantum state preparation~\cite{q_preparation, state_prep}, where rotational gates are applied to load the required amplitudes into the qubits. Note that an $N$-dimensional classical RSS vector can be encoded in $n = log(N)$-dimensional quantum register, which is an exponential saving is space.

As a simple example, a classical 2D RSS vector $\begin{bmatrix} \alpha \\ \beta \end{bmatrix}$ can be encoded into a single qubit as $\alpha \ket{0} + \beta \ket{1}$. This can be achieved by moving the state of a qubit in state $\ket{0}$ to be in state $\ket{\psi} = \alpha\ket{0} + \beta\ket{1}$ by rotating the qubit with angle $\theta$ around the Y-axis, where $\alpha = \cos{\frac{\theta}{2}}$ and $\beta = \sin{\frac{\theta}{2}}$. This can be achieved using the $R_y(\theta)$ gate for rotation around the Y-axis of the Bloch sphere~\cite{rygate}.


In a similar manner, the oracle $O_\phi$ is not only responsible for initializing the fingerprint register with the fingerprint data, it is also responsible for entangling the fingerprint data with the index so that it becomes in state $\frac{1}{\sqrt{M}}\sum_{j=0}^{M-1} \ket{\phi_j} \ket{j}$ as shown in Equation~\ref{eq:oracle_phi}.  The goal is to enter all fingerprint locations to the quantum circuit so that the similarity calculations can be performed in parallel. 
To do this we can use quantum state preparation~\cite{q_preparation, state_prep} to prepare the registers with $\frac{1}{\sqrt{M}}\begin{bmatrix} \alpha_{0,0} & .. & \alpha_{0,N-1} & .. & \alpha_{M-1,N-1}\end{bmatrix}^{T}$, where $M$ is the number of fingerprint locations, and $N$ is the number of BSs. Note also that a fingerprint with data from $N$ BS's at $M$ locations can be stored in a quantum registers of size $n+m$, where $n=log(N)$ and $m=log(M)$, as compared to the classical fingerprint size of $N \times M$.

\subsection{Example}
In this section, we give a simple numerical example of the proposed quantum circuit that is used to get the cosine similarity between the UE's RSS sample vector ($\psi$) and a fingerprint at two different locations ($\phi_j$, $j \in \{0,1\}$, $M$=2). Each RSS vector has the RSS from two different BS's ($N$=2). All vectors are unit vectors. 

Figure~\ref{fig:example_circuit} shows the quantum circuit used to obtain the cosine similarity between the online UE's RSS sample vector $\psi$ = $ \begin{bmatrix} 0.899 & 0.437 \end{bmatrix}$ (encoded in a quantum register $\ket{\psi}$), and each fingerprint vector $\phi_j$ at each index $j$, where $\phi_0 = \begin{bmatrix} 0.800 & 0.599 \end{bmatrix}$ and $\phi_1 = \begin{bmatrix} 0.543 & 0.839 \end{bmatrix}$ (encoded in a quantum register $\ket{\phi}$). 



Oracle $O_\psi$ is implemented using a rotational gate around the Y-axis ($R_y$) to set the UE's RSS sample register state to be $\ket{\psi}=\begin{bmatrix} 0.899 \\ 0.437 \end{bmatrix}$. This is done by applying a rotation with angle $\theta_\psi=2\times\arctan(\frac{b}{a})$, where $a$,$b$ are the required probability amplitudes of the test sample $\ket{\psi}$, i.e. $a=0.899$ and $b=0.437$, which results in $\theta_\psi=0.905$.

Oracle $O_\phi$ is implemented with the same idea as Oracle $O_\psi$, but here we are initializing the fingerprint register depending on the index value. So to initialize the fingerprint register with the first training sample $\phi_0=[0.800, 0.599]$, we need to apply an $R_y$ gate with $\theta_{\phi_0}=2\times\arctan(\frac{0.599}{0.800})=1.285$, and to initialize it with the second training sample $\phi_1=[0.543, 0.839]$, we need to apply another $R_y$ gate with $\theta_{\phi_1}=2\times\arctan(\frac{0.839}{0.543})=1.992$. One way to do that is by finding the difference between the two rotations $\textrm{diff}=(\theta_{\phi_0}-\theta_{\phi_1})$ and then applying the rotation in two steps, first we rotate the register with $\theta_{\phi_0}-\frac{\textrm{diff}}{2} = 1.638$; then if the index register has value $0$, we add this difference again by applying rotation with angle $\frac{\textrm{diff}}{2}=-0.353$; and if it has value $1$, we should apply the same rotation but in the opposite direction to reach the state $\bra{\phi_1}=[0.543, 0.839]$. We use $\textrm{CNOT}$ to flip the rotation direction controlled on the index register as shown in the initialization block in Figure~\ref{fig:example_circuit}. 

We ran this circuit with number of shots equals 1024 ($K=1024$), and we counted the cases where index register equals 0 (i.e. $\text{count}(i=0)$) and the cases where it equals 1 (i.e. $\text{count}(i=1)$), and for each case we counted the number of cases where ancilla qubit equals 0 (i.e. $\text{count}(a=0 \cap i=j)$). 
We measured 
$\text{count}(a=0 \cap i=0)=502$, and $\text{count}(a=0 \cap i=1)=436$. Since $\text{count}(a=0 \cap i=0) > \text{count}(a=0 \cap i=1)$, then we can estimate the UE's location as the location of $\phi_0$ stored in the fingerprint.

\section{Evaluation}
\label{sec:evaluation}

In this section, we implement our quantum algorithm in a \textit{\textbf{real}} testbed and evaluate its performance side-by-side against its classical counterpart and the state-of-the-art quantum positioning algorithms.

We start by describing our  testbed. Then, we show the accuracy using a real IBM quantum machine in a small testbed followed by  a larger scale experiment on the IBM quantum  simulator. After that, we quantify the theoretical space and time complexity of the proposed algorithm. Finally, we present experiments to evaluate different practical aspects of the algorithm.

We end the section with a discussion of the different aspects of the algorithm.


\subsection{Testbed Setup}
\begin{figure}[!t]
	\centerline
	{\includegraphics[width=0.4\textwidth, height=0.5\linewidth]{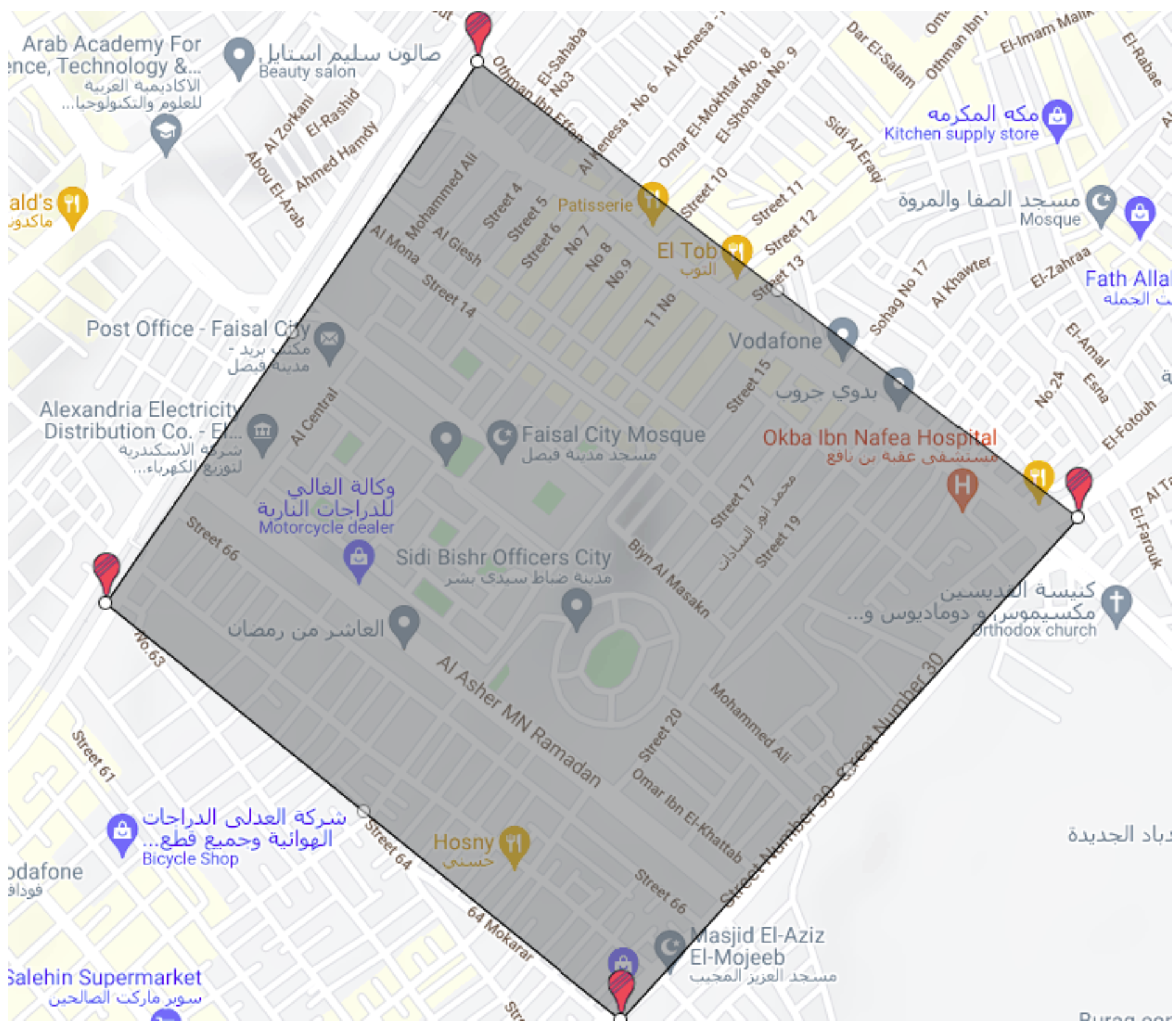}}
	\caption{Cellular outdoor testbed area.}
	\label{fig:testbed}
\end{figure}

We use a real cellular testbed in an  $0.2 \textrm{Km}^{2}$ urban area (Figure~\ref{fig:testbed}).  The area is covered by $21$ different cell-towers ($N=21$). 

In the offline phase, we use different Android devices to collect the fingerprint data at $44$ different locations uniformly distributed over the area of interest. Each device has a data collector software that collects GPS ground-truth locations, the base stations received signal strengths (RSS), and timestamps. We also collect an independent dataset of another $44$ samples to use as the online phase samples. In the online phase, the user's sample is sent to a quantum computer along with the fingerprint samples where the quantum circuit is run for $K$ times. Then we retrieve the results and calculate the fingerprint index with the maximum similarity to the user's sample and return the location of this fingerprint sample as the estimated user location.

\subsection{Accuracy Evaluation}



We evaluate the performance of our algorithm on a \textit{\textbf{real}} 5-qubits quantum IBM machine as well as the IBM quantum machine simulator. For the real quantum machine, we used the \textit{ibmq\_manila} machine with 5 qubits. Given the limited number of available qubits, we can only process a small testbed  with two base stations ($N=2$) and four fingerprint locations ($M=4$): one qubit for the ancilla, one qubit for encoding the RSS from two base stations for the unknown testing location, three qubits for  encoding the four fingerprint locations with their indices. To do that, we selected the two base stations that are most commonly heard in the testbed area as well as four fingerprint locations uniformly spread over the area.  
For simulation, we use the IBM Quantum Machine Simulator with the total testbed samples.

\begin{figure}[!t]
	\centerline
	{\includegraphics[width=0.29\textwidth, angle=-90]{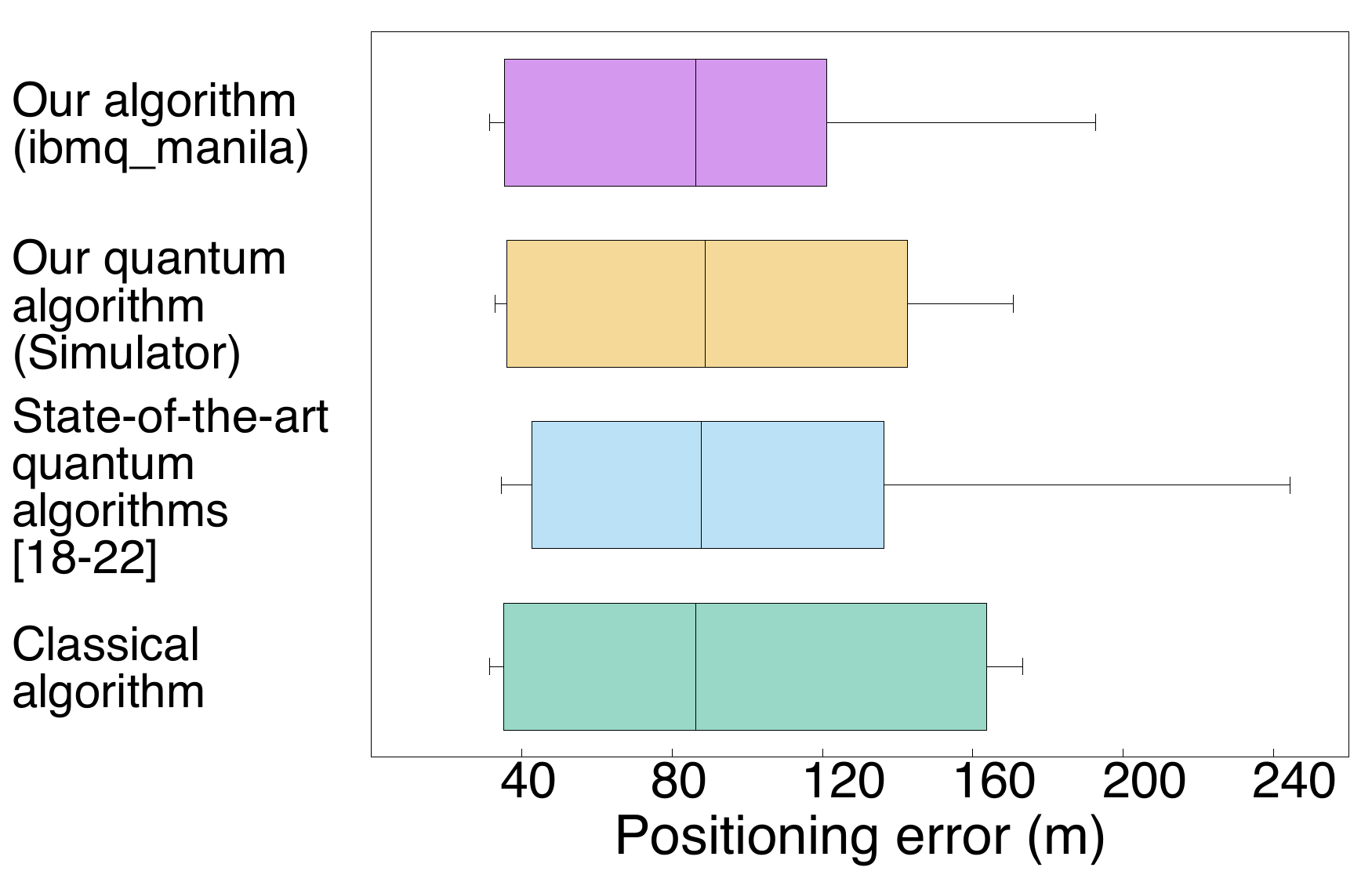}}
	\caption{ Real Quantum IBM machine experiment:  positioning error distributions comparing the proposed algorithm, to quantum simulator, state-of-the-art quantum algorithms~\cite{quantum_lcn, quantum_qce, quantum_vision, quantum_arx, device_indp_q}, and the classical algorithm.} 
	\label{fig:cdf_new_real_ibm}
\end{figure}

Figure~\ref{fig:cdf_new_real_ibm} shows a comparison between the positioning error distribution of our proposed quantum algorithm which runs in $o(\log(MN))$ implemented on \textit{ibmq\_manila} real machine and simulator, the state-of-the-art quantum positioning algorithms~\cite{quantum_lcn, quantum_qce, quantum_vision, quantum_arx, device_indp_q} which runs in $o(M\log(N))$, and the classical version of the cosine-similarity positioning algorithm which runs in $o(MN)$. The figure confirms that the proposed quantum algorithm can achieve the same accuracy as the classical counterpart and state-of-the-art quantum algorithms. This comes with the exponential enhancement in space and time in both $N$ and $M$ compared to the classical version. Moreover, it comes with the exponential gain in $M$ over the state-of-the-art quantum algorithms~\cite{quantum_lcn, quantum_qce, quantum_vision, quantum_arx, device_indp_q} as we quantify in Section~\ref{sec:complexity}.  The figure further \textit{\textbf{validates}} that the performance obtained from the simulator matches the performance of the real quantum machine, which we discuss next.



\begin{figure}[!t]
	\centerline
	{\includegraphics[width=0.26\textwidth, angle=-90]{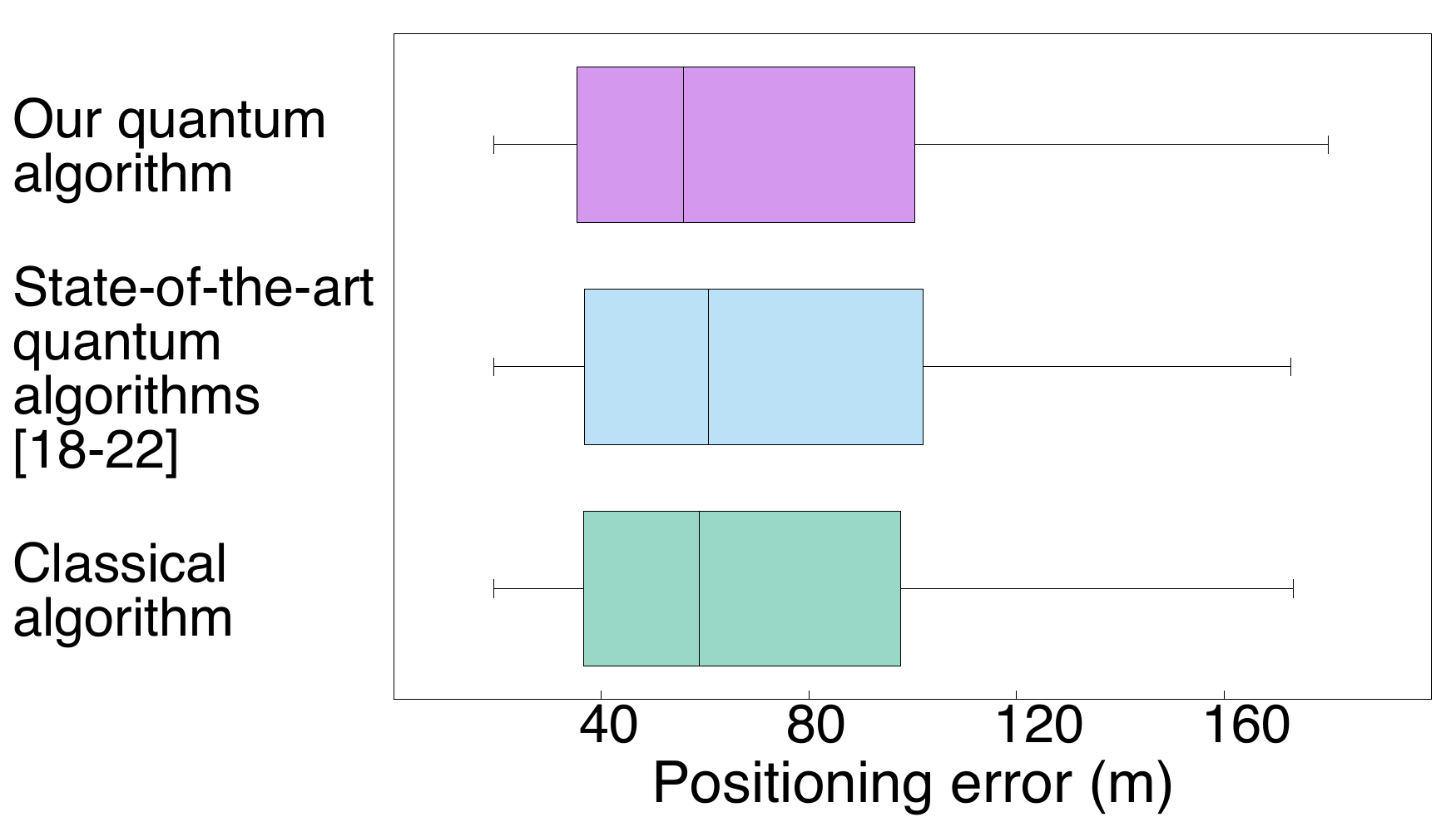}}
	\caption{IBM Quantum Simulator: positioning error distributions comparison on a larger testbed for the different algorithms.}
	\label{fig:cdf_new}
\end{figure}

To show the scalability of the algorithm, in the rest of this section we used the total testbed samples and implemented it over the IBM quantum machine simulator. Figure~\ref{fig:cdf_new} shows the positioning error distribution for our proposed quantum algorithm, the state-of-the-art quantum positioning \cite{quantum_lcn, quantum_qce, quantum_vision, quantum_arx, device_indp_q}, and the classical algorithm. The figure confirms that  our algorithm has the same accuracy as the state-of-the-art quantum algorithms and the classical version over the larger testbed. 

\begin{figure}[!t]
	\centerline
	{\includegraphics[width=0.45\textwidth]{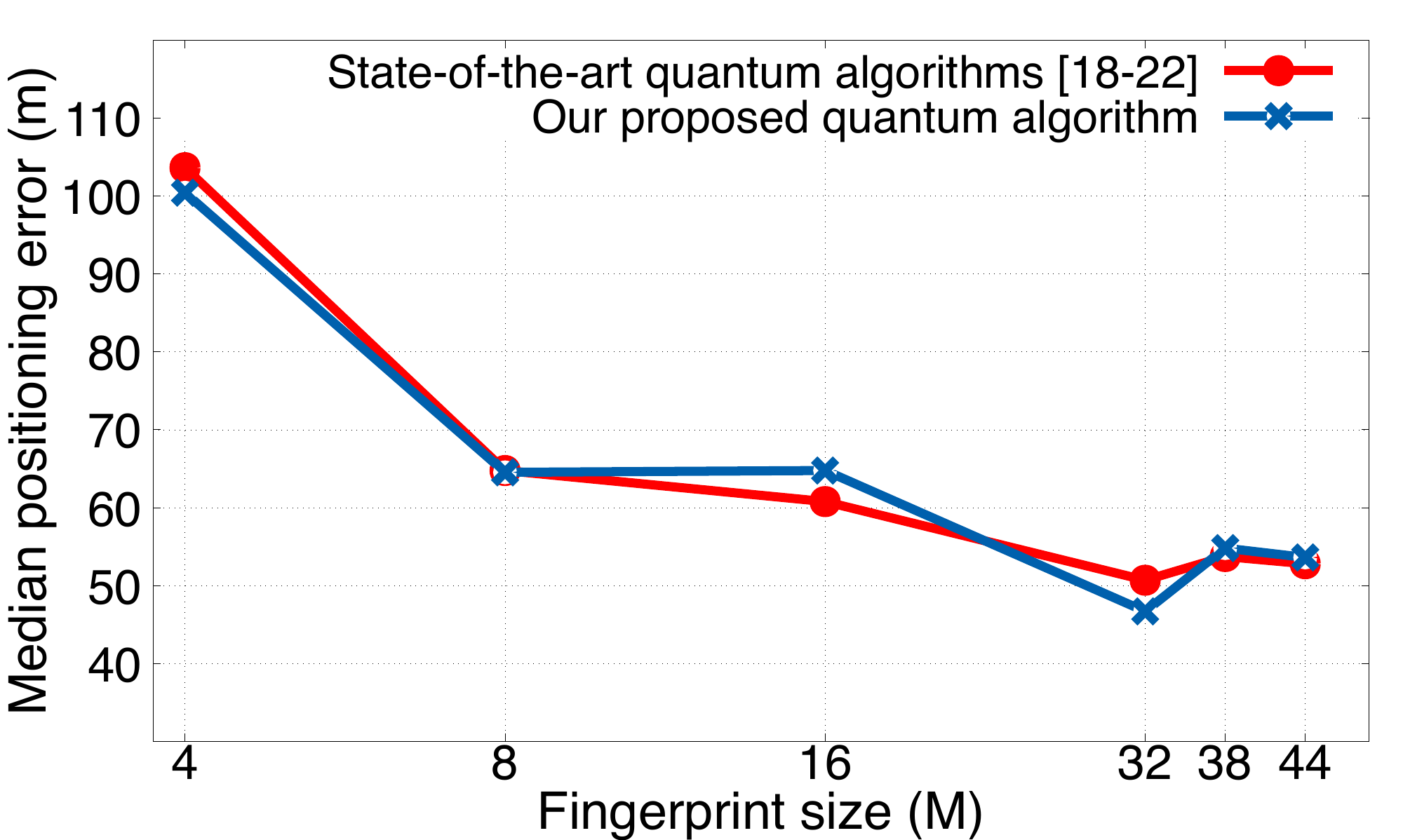}}
	\caption{Median positioning error for different numbers of fingerprint locations.} 
	\label{fig:total_M_bottom}
\end{figure}

Figure~\ref{fig:total_M_bottom} further compares the proposed quantum algorithm with the state-of-the-art  quantum algorithms 
 while increasing the number of fingerprint locations. The figure shows that, as expected, the higher the density of the fingerprint locations, the higher the accuracy will be. 
The figure also confirms that the proposed quantum algorithm can achieve the same accuracy as the $o(M \log(N))$ algorithms while having an \textbf{exponential saving} in the number of shots ($K$) needed to run the circuit as we show in the next section. 

\begin{figure}[!t]
	\centerline
	{\includegraphics[width=0.45\textwidth]{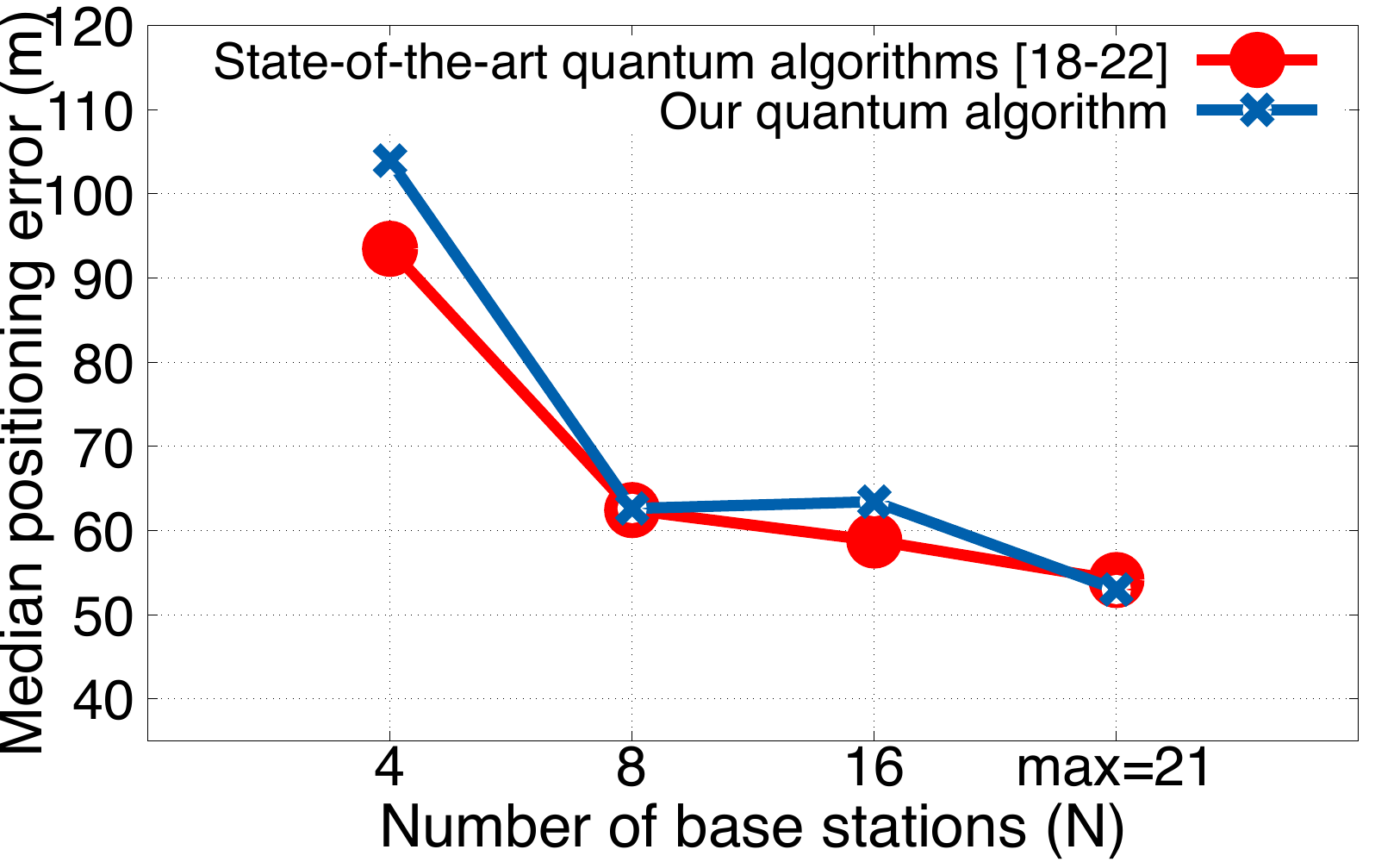}}
	\caption{Median positioning error for different numbers of base stations (N).} 
	\label{fig:different_n}
\end{figure}

Figure~\ref{fig:different_n}, shows the median positioning error for the proposed quantum algorithm with the state-of-the-art algorithms at different numbers of base stations ($N$). The figure highlights that the higher the number of base stations used in the positioning, the better the accuracy will be. It shows also that our algorithm gives similar results as the state-of-the-art at different numbers of base stations but with the exponential time and space enhancement.

\begin{figure}[!t]
	\centerline
	{\includegraphics[width=0.48\textwidth]{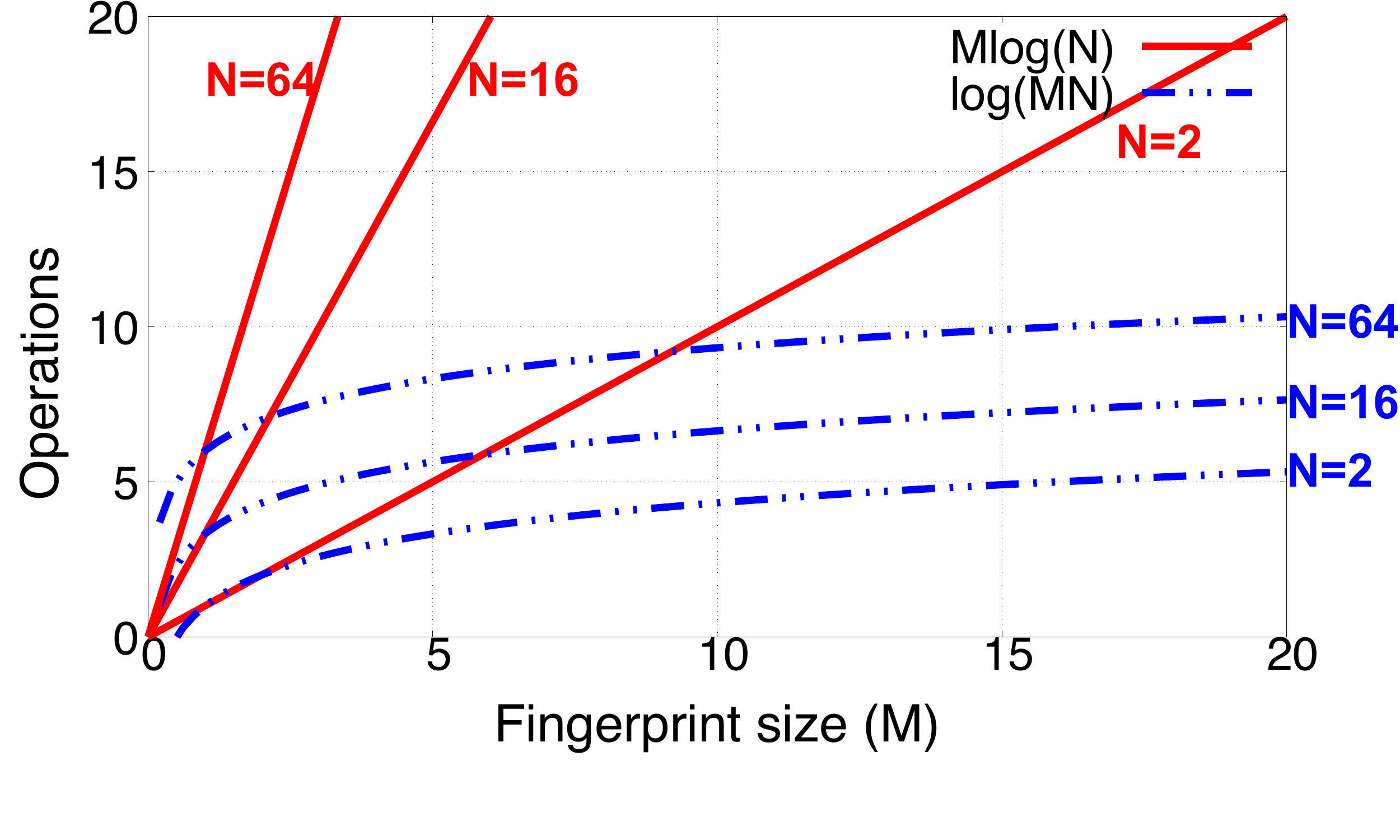}}
	\caption{The complexity of $o(M\log(N))$ algorithms~\cite{quantum_lcn, quantum_qce, device_indp_q, quantum_arx, quantum_vision} and our proposed $o(\log(MN))$ quantum algorithms as a function of the fingerprint size ($M$) at different number of BSs ($N$).}
	\label{fig:geogebra_tc}
\end{figure}

\begin{figure}[!t]
	\centerline
	{\includegraphics[width=0.45\textwidth]{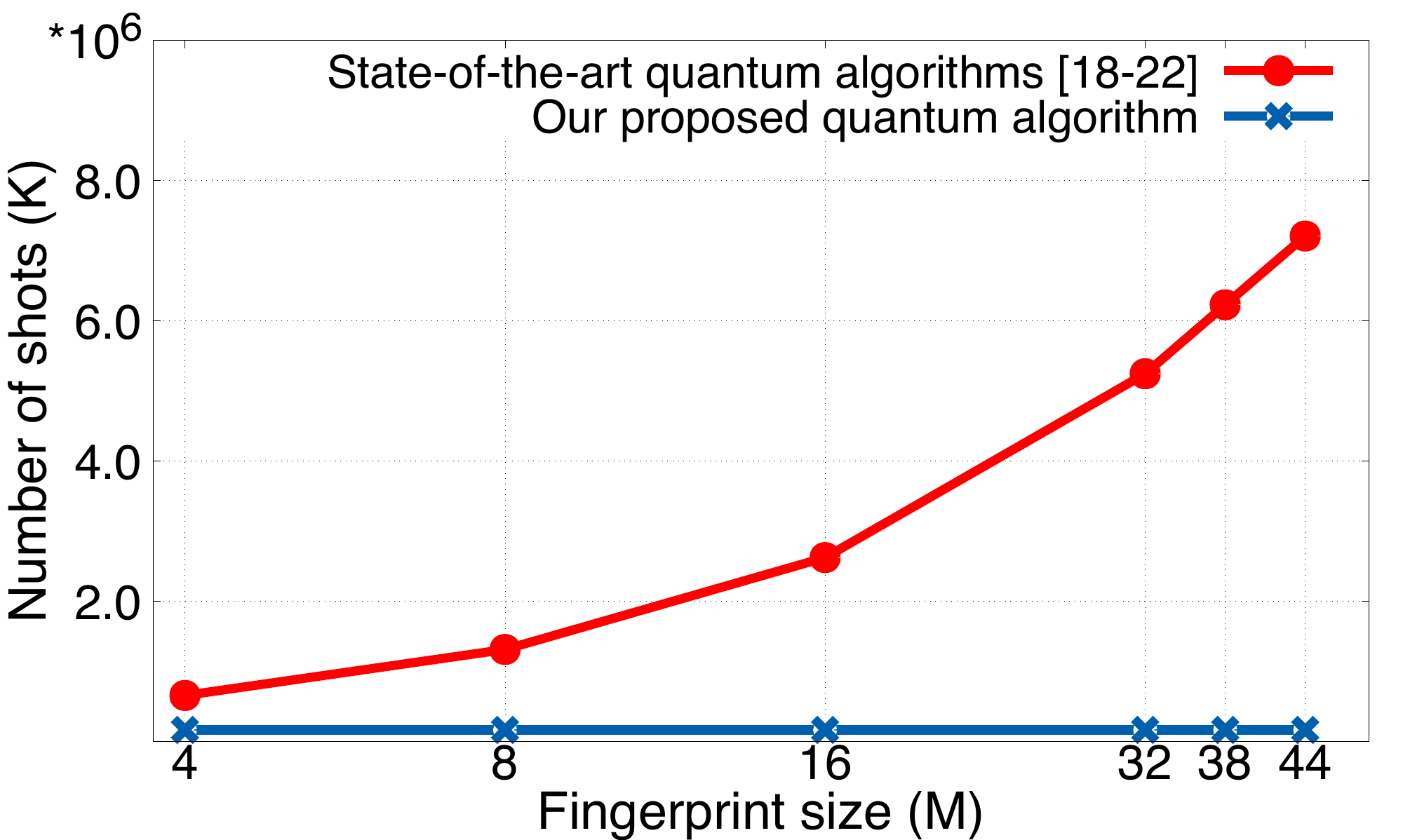}}
	\caption{Number of required shots ($K$) for different numbers of fingerprint locations.} 
	\label{fig:total_M_top}
\end{figure}

\begin{figure}[!t]
	\centerline
	{\includegraphics[width=0.4\textwidth]{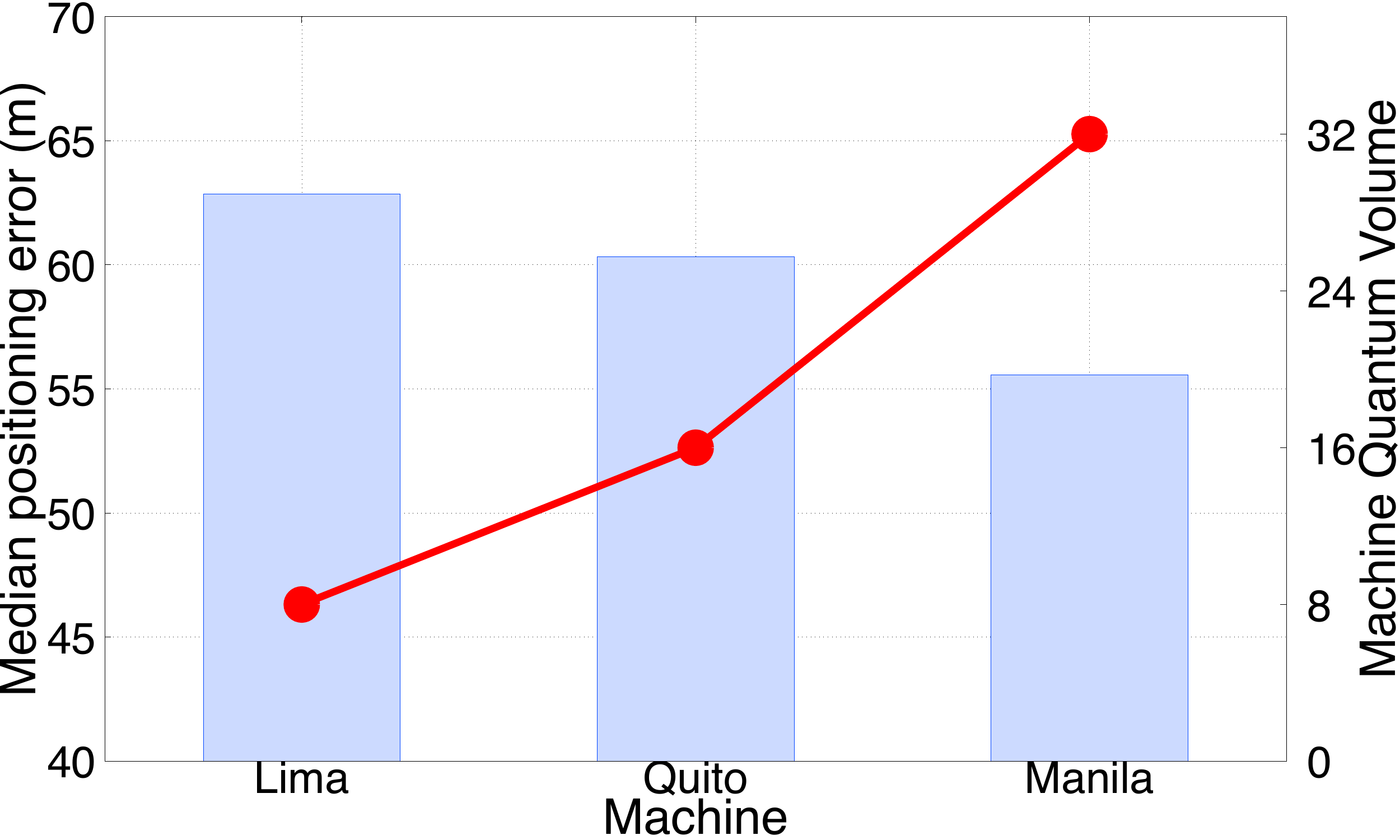}}
	\caption{Effect of the quantum volume (QV) of different quantum machines on the positioning accuracy.} 
	\label{fig:noise_compariosn}
\end{figure}

\begin{figure}[!t]
	\centerline
	{\includegraphics[width=0.45\textwidth]{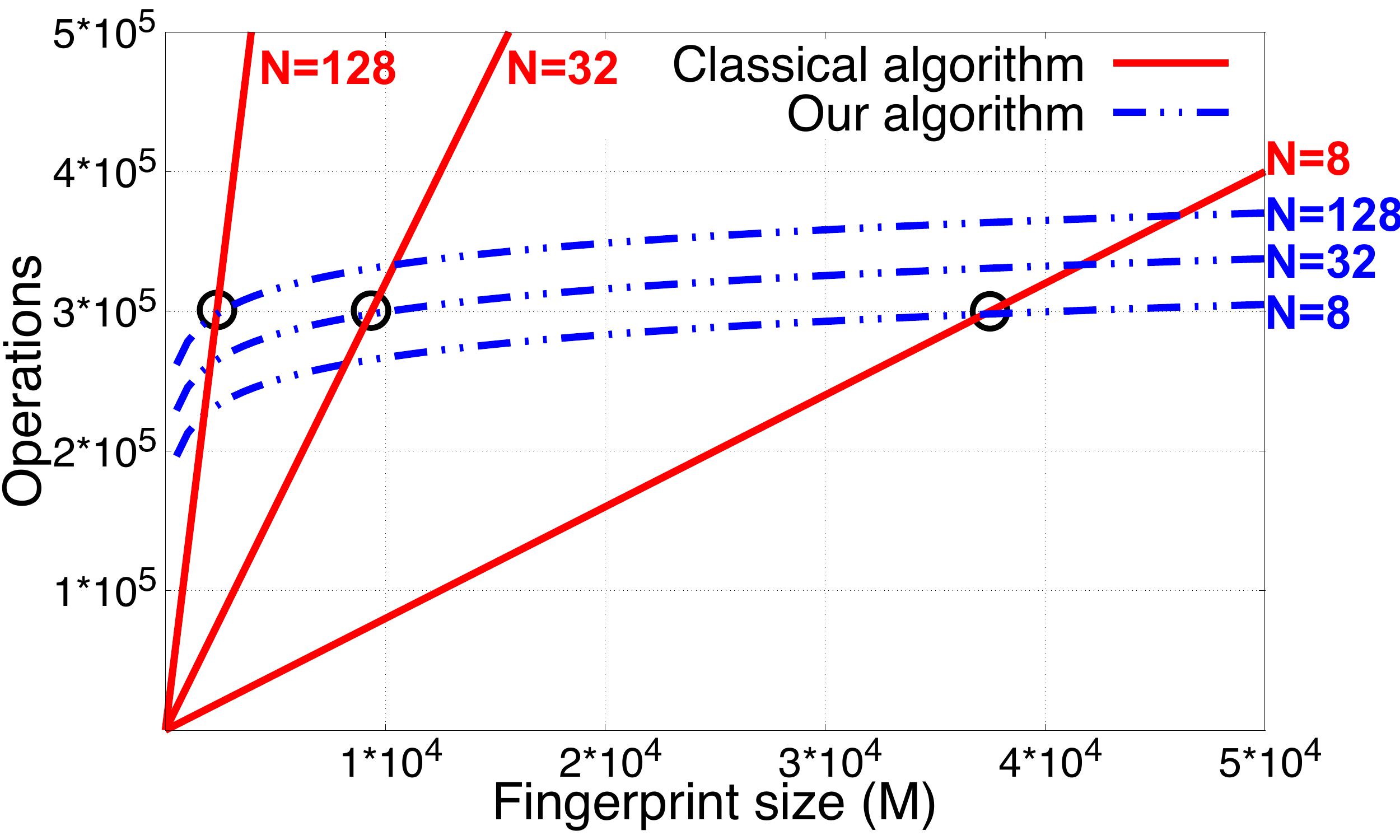}}
	\caption{Number of operations required for our quantum algorithm at $K=2^{14}$ vs the classical algorithm at different values of base stations $N$.} 
	\label{fig:class_vs_quantum}
\end{figure}



\subsection{Complexity Analysis}
\label{sec:complexity}
%
%
%
%
%
In general, the complexity of the fingerprint-based techniques depends on the number of fingerprint locations ($M$) and the number of BSs ($N$).

The first stage of the proposed algorithm, the Initialization Stage, uses quantum state preparation techniques to initialize the quantum registers with the classical user sample and fingerprint data. Efficient state preparation techniques can be used for state preparation to load the data in logarithmic complexity, such as Quantum Random Access Memory (QRAM)\cite{qram1, qram2, state_prep}, where a vector with size $N$ can be loaded in $o(\log(N))$~\cite{qram1} in parallel into a qubit register and conditional rotations are performed to encode the vector data as amplitudes in the quantum registers. Therefore, loading $M$ fingerprint vectors each with $N$ values will require $o(\log(MN))$ complexity. Moreover, quantum sensors~\cite{qsensor4, qsensor3, qsensor2, qsensor1} are evolving over time and the proposed algorithm can leverage this development by taking the data as input directly from quantum sensors,  which can make the quantum state preparation step complexity $o(1)$. 

The second stage is the quantum similarity matching, which is done using the Swap Test where two $H$ gates are used on the ancilla qubit, and $\log(N)$ CSWAP gates are used to entangle the ancilla qubit, with the UE's RSS register and the fingerprint register. This leads to $o(\log(N))$ complexity used in this stage.

Finally,  to find the fingerprint index with the maximum cosine similarity, we run the circuit for $K$ times and observe the circuit output to find the fingerprint index with the maximum count of measuring the ancilla qubit output as zero, i.e. finding index value $j$ where $count(a=0 \cap i=j)$ is maximum. The index register length is $\log(M)$, therefore, measuring the output of the register has a complexity of $o(\log(M))$.
Hence, the algorithm has an overall complexity of $o(\log(MN))$, compared to the state-of-the-art quantum algorithms~\cite{quantum_lcn, quantum_qce, quantum_vision, quantum_arx, device_indp_q} that take $o(M\log(N))$.

Note that the proposed quantum algorithm complexity of $o(\log(MN))$ is \textbf{more than exponentially better} than the other-state-of-the-art quantum algorithms complexity of $o(M\log(N))$, as an exponential enhancement would reduce the complexity to  $o(\log(M)\log(N))$ only.

Figure~\ref{fig:geogebra_tc} shows the difference between the complexity of the state-of-the-art quantum algorithms ($o(M\log(N))$) \cite{quantum_lcn, quantum_qce, quantum_vision, quantum_arx, device_indp_q} and our proposed quantum algorithm ($o(\log(MN))$) for different values of $N$ and $M$. The figure highlights that there is a significant better than-exponential gain of the proposed algorithm compared to the other state-of-the-art quantum algorithm. 

Figure~\ref{fig:total_M_top} shows the number of shots $K$ used for the state-of-the-art quantum algorithms \cite{quantum_arx, quantum_lcn, quantum_qce, quantum_vision, device_indp_q} and our proposed algorithm at different number of fingerprint sizes ($M$). The figure highlights that the proposed algorithm can have an exponential saving in the required number of circuit runs.

\subsection{Practical Considerations}
In this section, we evaluate some practical considerations of the proposed algorithm: the effect of machine noise on accuracy and the effect of the number of shots on  the complexity.

We use three quantum machines that have $5$ physical qubits each: \textit{ibmq\_manila}, \textit{ibmq\_quito}, and \textit{ibmq\_lima}. Each quantum machine has different characteristics that can affect the overall accuracy~\cite{decoherence, gates_error, q_volume}. To capture all these different characteristics in one metric, the quantum volume (QV) has been proposed~\cite{q_volume,q_volume_2}, which measures a quantum computer's performance taking into account gates' errors, measurement errors, quality of the circuit compiler, among others~\cite{qbible}. The higher the QV is, the less error-prone the machine is. 
Figure~\ref{fig:noise_compariosn} shows that, as expected, the localization accuracy increases as the quantum volume increases (i.e. the quantum machine's noise is lower).

Finally, Figure~\ref{fig:class_vs_quantum} compares the total number of operations required for our proposed quantum algorithm and its classical counterpart, considering number of shots $K=2^{14}=16384$ for the quantum algorithm at different numbers of base stations ($N$). The black circles show the point where the quantum algorithm has the same number of operations as the classical version. The figure highlights that the proposed quantum algorithm can perform much better than the classical algorithm at high fingerprint data size and high number of base stations, taking into account the number of shots required in the quantum algorithm.

\subsection{Discussion}
\label{sec:disc}


The cosine similarity-based \textbf{classical} positioning algorithm has a \textit{\textbf{quadratic}} complexity ($o(MN)$) in both space and time, where $M$ is the number of fingerprint locations and $N$ is the number of BSs in the environment. In contrast, the proposed quantum positioning algorithm has a \textit{\textbf{sub-linear}} complexity ($o( \log(MN))$) in space and time.

Unlike the state-of-the-art quantum algorithms that need sub-quadratic space and time ($o(M\log(N))$), the proposed quantum algorithm sub-linear complexity offers an exponential enhancement in the number of fingerprint locations for both space and running time. 

%
The exponential improvement in the number of fingerprint locations ($M$) enables us to get more accurate positioning using larger fingerprint data.  
 On the other hand, the exponential improvement in the number of BSs ($N$) enables us to build a fingerprint with a large-scale of heterogeneous BSs (e.g. cellular towers, WiFi APs, BLE), which has the potential of higher positioning accuracy. This can be used for different scenarios, e.g., where each device can be used as a reference point for positioning, as in intelligent transportation systems, connected and automated vehicles, and industrial internet of things (IIOT) applications. 
 All of these are potential targets for 5G/6G high-accuracy low-latency positioning as defined by 3GPP Release-17 \cite{3gpp_rel17}. 
 
 On the other hand, the storage space required for offline quantum fingerprint building is reduced exponentially, as the fingerprint size is reduced from  $o(MN)$ to $o(\log(MN))$.

%
With the emergence of quantum co-processors (similar to GPUs)~\cite{frangou2019first}, the proposed quantum algorithm can be completely run on the user equipment, e.g., for privacy issues. For this, the fingerprint data can be downloaded from a server to the UE. In such cases, the proposed quantum algorithm saves both the storage space required on the UE as well as the required download bandwidth.

We show experimental results for the algorithm implementation on a real quantum machine with $5$ qubits. However, more advanced quantum machines are available with a higher number of qubits reaching up to $433$ qubits as in the \textit{ibm\_seattle} machine~\cite{ibmq_res} (though not freely accessible). Given that the algorithm circuit needs $1 + \log(M) + 2\log(N)$ qubits, this machine can accommodate a very large number of fingerprint samples and base stations. 
 The quantum machines have different error characteristics due to different sources of noise such as the quantum decoherence, gates errors and readout error~\cite{decoherence, gates_error, q_volume}, and all need to be taken into consideration. However, The development of quantum computers is growing fast, allowing quantum algorithms to be more practical and feasible in the foreseeable future~\cite{q_compu_evolution}.
Furthermore, our results show that the number of shots $K$ affects the overall positioning accuracy. Although the number of shots $K$ is considered constant in the theoretical analysis, it should be taken into account in practical consideration since the required number of shots may be high.

\section{Conclusion}
\label{sec:conclusion}

In this paper, we have presented a cosine similarity-based quantum algorithm for enabling large-scale worldwide positioning. Unlike the classical techniques, which need $o(MN)$ time and space, the proposed quantum algorithm requires $o(\log(MN))$ time and space for a fingerprint with $M$ locations and $N$ BSs. We implemented the proposed algorithm on a real IBM quantum machine as well as a simulator and evaluated it in a real cellular outdoor testbed. We also compared the proposed algorithm with the state-of-the-art quantum algorithms for positioning, showed how the algorithm accuracy changes across different quantum machines with different noise profiles, quantified its complexity, and discussed its practicality. The proposed quantum algorithm can provide an exponential saving in both the number of fingerprint locations and the number of BSs, taking positioning systems a step toward a more accurate and ubiquitous positioning that can work on a worldwide scale and meet the requirements of the next generation 5G, 6G, and beyond.

Currently, we are working on multiple research directions for positioning systems using quantum computing including exploring different quantum similarity metrics, using quantum computing techniques for floor detection, among others.

\appendices
\section{Derivations of Equation~\ref{eq:gamma_tilde} and \ref{eq:p_a_given_i}}
\label{appendix:A}

\subsection{Derivation of Equation~\ref{eq:gamma_tilde}}
\label{appendix:A_1}

Before measuring the ancilla qubit and index register in circuit shown in Figure~\ref{fig:main_circuit}, the state of the quantum system is:

\begin{equation}\nonumber
\begin{aligned}
     \ket{\gamma_4} = \frac{1}{2\sqrt{M}} \sum_{j=0}^{M-1} (\ket{0}[\ket{\psi}\ket{\phi_j} + \ket{\phi_j}\ket{\psi}]\\
    + \ket{1}[\ket{\psi}\ket{\phi_j} - \ket{\phi_j}\ket{\psi}])\ket{j}
\end{aligned}
\end{equation}

which can be rewritten for simplicity as:
\begin{equation}\nonumber
\begin{aligned}
     \ket{\gamma_4} = \frac{1}{2\sqrt{M}} \sum_{j=0}^{M-1} \ket{\zeta_{j}}\ket{j}
\end{aligned}
\end{equation}
where $\ket{\zeta_{j}}=\ket{0}[\ket{\psi}\ket{\phi_j} + \ket{\phi_j}\ket{\psi}]+ \ket{1}[\ket{\psi}\ket{\phi_j} - \ket{\phi_j}\ket{\psi}] = \ket{0\psi\phi_j} + \ket{0\phi_j\psi} + \ket{1\psi\phi_j} - \ket{1\phi_j\psi}$. \\

After measuring index register, the quantum system moves to the normalized state:

$\frac{1}{\lVert \zeta_{j} \rVert}\ket{\zeta_{j}}$, where $\lVert \zeta_{j} \rVert$ is the euclidean norm of $\ket{\zeta_{j}}$. The value of ${\lVert \zeta_{j} \rVert}$ is calculated as follows:

\begin{gather*}
    {\lVert \zeta_{j} \rVert}^2=\braket{\zeta_j|\zeta_j}\\ 
    = \braket{0\psi\phi_j|0\psi\phi_j} + \braket{0\psi\phi_j|0\phi_j\psi} + \braket{0\psi\phi_j|1\psi\phi_j} - \braket{0\psi\phi_j|1\phi_j\psi}\\
    + \braket{0\phi_j\psi|0\psi\phi_j} + \braket{0\phi_j\psi|0\phi_j\psi} + \braket{0\phi_j\psi|1\psi\phi_j} - \braket{0\phi_j\psi|1\phi_j\psi}\\
    + \braket{1\psi\phi_j|0\psi\phi_j} + \braket{1\psi\phi_j|0\phi_j\psi} + \braket{1\psi\phi_j|1\psi\phi_j} - \braket{1\psi\phi_j|1\phi_j\psi}\\
    - \braket{1\phi_j\psi|0\psi\phi_j} - \braket{1\phi_j\psi|0\phi_j\psi} - \braket{1\phi_j\psi|1\psi\phi_j} + \braket{1\phi_j\psi|1\phi_j\psi}
\end{gather*}
And since the dot product of normalized vector by itself equals 1, and the dot product of orthogonal vectors $\ket{0}$ and $\ket{1}$ equals 0, we can say that:
\begin{gather*}
    {\lVert \zeta_{j} \rVert}^2 \\
    = 1 + \braket{0\psi\phi_j|0\phi_j\psi} + 0 - 0
    + \braket{0\phi_j\psi|0\psi\phi_j} + 1 + 0 - 0\\
    + 0 + 0 + 1 - \braket{1\psi\phi_j|1\phi_j\psi}
    - 0 - 0 - \braket{1\phi_j\psi|1\psi\phi_j} + 1
\end{gather*}

And it can be proved mathematically that $\braket{0\psi\phi_j|0\phi_j\psi} = \braket{0\phi_j\psi|0\psi\phi_j} = \braket{1\psi\phi_j|1\phi_j\psi} = \braket{1\phi_j\psi|1\psi\phi_j}$, hence we get ${\lVert \zeta_{j} \rVert} = 2$. Now we can write the quantum system state after measuring index register as in Equation~\ref{eq:gamma_tilde}:

\begin{equation}\nonumber
\frac{1}{2}\ket{\zeta_{j}} = \frac{1}{2}(\ket{0}[\ket{\psi}\ket{\phi_j} + \ket{\phi_j}\ket{\psi}] + \ket{1}[\ket{\psi}\ket{\phi_j} - \ket{\phi_j}\ket{\psi}])  
\end{equation}

\subsection{Derivation of Equation~\ref{eq:p_a_given_i}}
\label{appendix:A_2}
To find the conditional probability $p(a=0 | i=j)$ we need to find the probability that the ancilla qubit equals zero. For simplicity, we can write the state obtained in Equation~\ref{eq:gamma_tilde} as:

\begin{gather*}\nonumber
    \frac{1}{2}(\ket{0}[\ket{\psi}\ket{\phi_j} + \ket{\phi_j}\ket{\psi}] + \ket{1}[\ket{\psi}\ket{\phi_j} - \ket{\phi_j}\ket{\psi}]) \\
    = \frac{1}{2}(\lVert \eta_0 \rVert \ket{0}\frac{\ket{\eta_0}}{\lVert \eta_0 \rVert} + \lVert \eta_1 \rVert\ket{1}\frac{\ket{\eta_1}}{\lVert \eta_1 \rVert})
\end{gather*}

where $\ket{\eta_0}=[\ket{\psi}\ket{\phi_j} + \ket{\phi_j}\ket{\psi}]$ and $\ket{\eta_1}\\=[\ket{\psi}\ket{\phi_j} - \ket{\phi_j}\ket{\psi}]$, and $\lVert \eta_{i} \rVert$ is the euclidean norm of $\ket{\eta_{i}}$. The probability that the ancilla qubit is in state $\ket{0}$ equals $p(a=0|i=j) = \big(\frac{\lVert \eta_0 \rVert}{2}\big)^2$. Note that this is after measuring index register (conditional probability).


The value of $\lVert \eta_0 \rVert$ is found as follows: 

\begin{gather*}
    {\lVert \eta_0 \rVert}^2=\braket{\eta_0|\eta_0}\\ 
    = \braket{\psi\phi_j|\psi\phi_j} +  \braket{\psi\phi_j|\phi_j\psi} + \braket{\phi_j\psi|\psi\phi_j} +  \braket{\phi_j\psi|\phi_j\psi}\\
    = 2 + 2\braket{\psi\phi_j|\phi_j\psi} = 2+2\big|\braket{\psi|\phi_j}\big|^2
\end{gather*}

So the probability of $a=0$ given that $i=j$ is (as shown in Equation~\ref{eq:p_a_given_i}):
\begin{gather*}
    p(a=0|i=j) = (\frac{\lVert \eta_0 \rVert}{2})^2 = \frac{\lVert \eta_0 \rVert^2}{4}\\ 
    = \frac{1}{4}(2+2\big|\braket{\psi|\phi_j}\big|^2)  = \frac{1}{2} + \frac{1}{2} \big|\braket{\psi|\phi_j}\big|^2
\end{gather*}


\section{Proof that $cos(\psi,\phi_j) \propto \text{count}(a=0 \cap i=j)$}
\label{appendix:B}

Since the cosine similarity is directly proportional to $p(a=0|i=j)$ (from Equation~\ref{eq:cos_final}). And since $p(a=0|i=j)=\frac{\text{count}(a=0 \cap i=j)}{\text{count}(i=j)}$ (from Equation~\ref{eq:probability_by_counting}), then we can say that the cosine similarity is directly proportional to $\text{count}(a=0 \cap i=j)$ if $\text{count}(i=j)$ is equal for all $j \in \{0,..,M-1\}$.

To prove that $\text{count}(i=j)$ is equal for all $j \in \{0,..,M-1\}$, we start from Equation~\ref{eq:gamma_bar}, where the quantum system is in the state:
\begin{equation}\nonumber
\begin{aligned}
     \ket{\gamma_4} = \frac{1}{2\sqrt{M}} \sum_{j=0}^{M-1} \ket{\zeta_{j}}\ket{j}
\end{aligned}
\end{equation}
where $\ket{\zeta_{j}}=\ket{0}[\ket{\psi}\ket{\phi_j} + \ket{\phi_j}\ket{\psi}]+ \ket{1}[\ket{\psi}\ket{\phi_j} - \ket{\phi_j}\ket{\psi}] = \ket{0\psi\phi_j} + \ket{0\phi_j\psi} + \ket{1\psi\phi_j} - \ket{1\phi_j\psi}$. \\

As shown in Appendix A, the euclidean norm of $\ket{\zeta_{j}}$ is  ${\lVert \zeta_{j} \rVert} = 2$. Therefore, the system is in the normalized state:
\begin{equation}\nonumber
\begin{aligned}
     \ket{\gamma_4} = \frac{1}{\sqrt{M}} \sum_{j=0}^{M-1} (\frac{1}{2}\ket{\zeta_{j}})\ket{j}
\end{aligned}
\end{equation}
which means that the probability of measuring the index register is $p(i=j)=\frac{1}{M}$ for all values of $j \in {\{0,..,M-1}\}$, i.e.  $\text{count}(i=j)$ is equal for all $j \in \{0,..,M-1\}$, which proves that $ cos(\psi,\phi_j)\propto \text{count}(a=0 \cap i=j)$, i.e. $\argmax_j (cos(\psi,\phi_j)) = \argmax_j (\text{count}(a=0 \cap i=j))$. 

\bibliographystyle{unsrt}

\begin{IEEEbiography}
[{\includegraphics[width=1in,height=1in,clip,keepaspectratio]{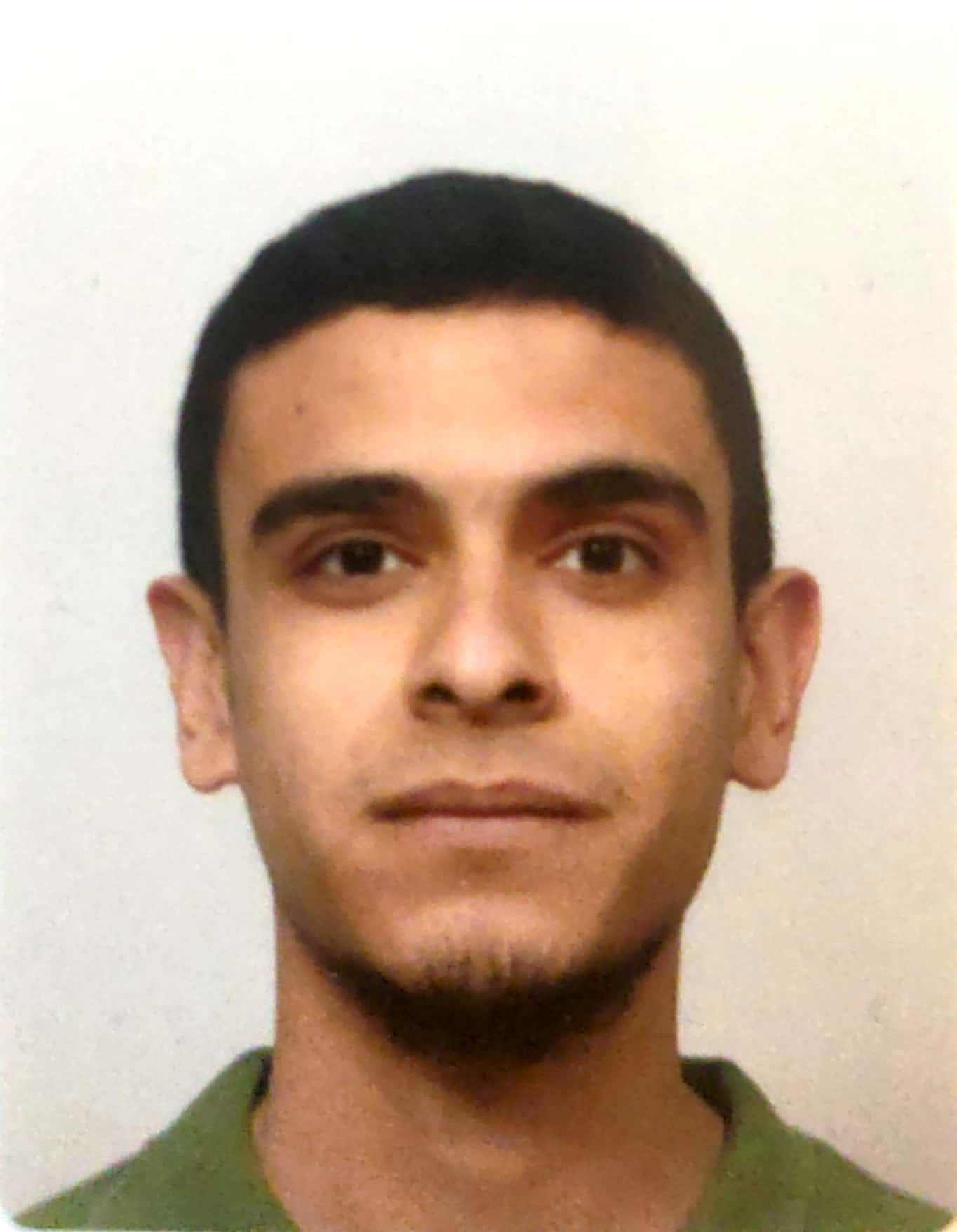}}]{Yousef Zook}
is a M.Sc. student at the Computer and Systems Engineering department at  Alexandria University, Egypt. He got his Bachelor from the same department in 2019. His research interests include localization systems and quantum computing.
\end{IEEEbiography}

\begin{IEEEbiography}
[{\includegraphics[width=1in,height=2in,clip,keepaspectratio]{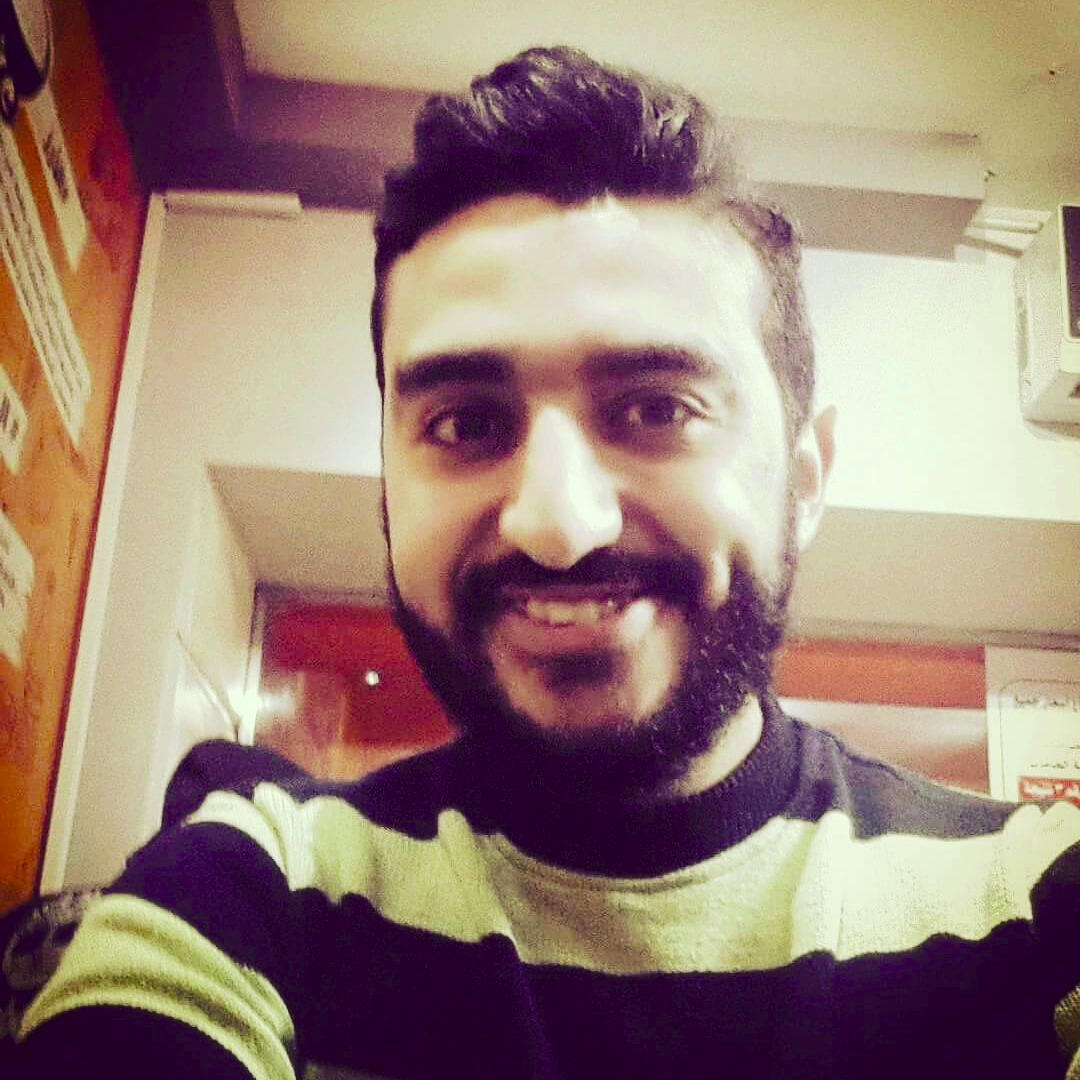}}]{Ahmed Shokry}
is a research assistant at the American University in Cairo, Egypt. 
Ahmed has received his B.Sc. in Computer and Systems Engineering from Alexandria University, Egypt in 2013. He also received his M.Sc. in Computer Science and Engineering from the same university in 2019. His research interests include applied machine/deep learning, location determination systems, quantum algorithms, and theoretical computer science.
\end{IEEEbiography}

\begin{IEEEbiography}    [{\includegraphics[width=1in,height=2in,clip,keepaspectratio]{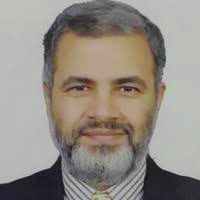}}]{Moustafa Youssef} 
 is a professor at the American University in Cairo and the University of New South Wales, Australia. He is the founder \& director of the Wireless Research Center of Excellence, Egypt. His research interests include mobile wireless networks, mobile computing, location determination technologies, pervasive computing, and quantum computing. He is an Associate Editor for IEEE TMC and ACM TSAS, served as the Lead Guest Editor of the IEEE Computer Special Issue on Transformative Technologies and an Area Editor of ACM MC2R as well as on the organizing and technical committees of numerous prestigious conferences. He is the recipient of the 2003 University of Maryland Invention of the Year award, the 2010 TWAS-AAS-Microsoft Award for Young Scientists, the 2013 and 2014 COMESA Innovation Award, multiple Google Research Awards, among many others. He is also an IEEE and ACM Fellow.
\end{IEEEbiography}

\end{document}